\documentclass[12pt, draftclsnofoot, onecolumn]{IEEEtran}
 \usepackage{cite}
 \usepackage{graphicx}
 \usepackage{bm}
 \usepackage{textcomp}
 \usepackage{amssymb}
 \usepackage{amsmath}
 \usepackage{lipsum}
 \usepackage{subfigure}

 \usepackage{epstopdf}
 \usepackage{color}
 \DeclareMathOperator*{\argmin}{\arg\!\min}
 
\newtheorem{my_theorem}{Theorem} 

\newtheorem{my_lemma}{Lemma}

   \newtheorem{my_corr}{Corollary}

 \usepackage{enumitem}
 \makeatletter
 \def\BState{\State\hskip-\ALG@thistlm}
 \makeatother
 \IEEEoverridecommandlockouts
\begin{document}
	\title{Energy Efficiency of Opportunistic Device-to-Device Relaying Under Lognormal Shadowing}
	
	\author{  S.~M.~ Zafaruddin,~\IEEEmembership{Member,~IEEE,} Jan Plachy,~\IEEEmembership{Student Member,~IEEE,}  Zdenek Becvar,~\IEEEmembership{Senior Member,~IEEE,} and Amir Leshem,~\IEEEmembership{Senior Member,~IEEE}
		
		\thanks{ S.~M.~ Zafaruddin and Amir Leshem are with Faculty of Engineering, Bar-Ilan University, 	Ramat Gan 52900, Israel (email:\{smzafar, leshema\}@biu.ac.il ).  Jan Plachy and  Zdenek Becvar are with Dpt. of Telecommunication Eng., Faculty of Electrical Engineering, Czech Technical University in Prague, Czech Republic (email:\{jan.plachy, zdenek.becvar\}@fel.cvut.cz). This work was supported by Grant No. 8G15008 funded by MEYS in the framework of the Czech-Israel project  MSMT-10795/2015-1 and by the Israeli Ministry of Science and Technology under grant 3-13038 for cooperation with the Czech Republic, and by  ISF grant 903/2013. S.~M.~ Zafaruddin was partially funded by the Israeli Planning and Budget Committee (PBC) post-doctoral fellowship. }
	}
	\maketitle

\begin{abstract}
Energy consumption is a major limitation of low power and mobile  devices. Efficient transmission protocols are required to minimize an energy consumption of  the mobile devices for ubiquitous connectivity in the next generation  wireless networks. Opportunistic schemes  select  a single relay using the criteria of the best channel and  achieve a near-optimal  diversity performance in a cooperative wireless system.
 In this paper, we study the energy efficiency of the opportunistic schemes for device-to-device communication. In the opportunistic approach, an energy consumed by devices is minimized by  selecting a single neighboring device as a relay using the criteria of minimum consumed energy in each transmission in the uplink of a wireless network.    We derive analytical bounds and  scaling laws on the expected  energy consumption when the devices experience log-normal shadowing   with respect to a base station considering both the transmission as well as circuit energy consumptions. We show that the protocol improves the energy efficiency of the network comparing to the direct transmission even if only a few devices are considered for relaying.  We also demonstrate the effectiveness of the protocol by means of simulations  in realistic scenarios of the wireless network.
\end{abstract}

\section{Introduction}
  The upcoming 5G technology is expected to deliver  high-speed data transmission  in order of tens of Gbps \cite{Andrews2014}, \cite{Shafi2017}. This will improve the quality of service of existing multimedia applications and will pave the way for more ambitious applications such as Tactical Internet, Internet of Things (IoT), Internet of vehicles (IoV),  Machine to Machine (M2M) communications, and e-healthcare \cite{Aigwal2016}.  These rate-demanding services  are highly energy-consuming and would increase run-time usage of devices. The  devices are equipped with batteries of a limited capacity, which can quickly run down if the energy consumption is not addressed properly. Efficient transmission protocols are needed to reduce  the energy consumption of mobile devices  since improvement in battery capacity is unable to keep pace with increasing power dissipation of signal processing and transmission circuits.   In addition to  conventional quality of service parameters, such as spectral efficiency and latency, also energy efficiency (measured in bits/Joule)  has become an important feature of wireless networks \cite{Li2011}, \cite{Freng2013}, \cite{Buzzi2015_survey_JSAC}.

Optimized  allocation of resources (typically bandwidth and power)  provides substantial energy efficiency  gains, however, at the cost of  throughput reduction for various wireless  systems \cite{Buzzi2008, Miao2011, Zappone2016jsac, Zappone2016wsa, Zappone2016siam}.  To improve the energy efficiency of wireless networks, cooperative relaying techniques have been investigated (see, e.g., \cite{Sun2013, Zappone2014, Cheung2013}) and energy-efficient multi-relay selection schemes with power control have been proposed (see, e.g., \cite{Himsoon2007_JSAC, Madan2008, Lim2013}). However, the energy efficiency in cooperative communication can degrade as the number of cooperators rises \cite{Zhou20082}.  In this paper, we focus on analyzing a  protocol with a single-relay selection that minimizes the transmission energy and increase the energy efficiency of all devices in a wireless network.

The single relay-based cooperative transmission provides significant performance improvement  by exploiting diversity from  spatially distributed nodes in a wireless network \cite{Bletsas2006, Krikidis2008, Jing2009,Ikki2010,Kalantari2011,Nosratinia2004,Li2012}.  The criteria for relay selection is based on the quality of channel or the signal to noise ratio (SNR) at the relays.  This scheme is very popular when attempting to minimize transmission energy and maximize the lifetime of wireless sensor networks in particular \cite{Chen2005_CL, Zhao2005, chen2007transmission, chen2007integrated, Cohen_2010_TSP, Huang2008, Zhou2008}.  It is noted that the state of the art cooperative schemes select a single device from the whole network which increases overhead energy  and latency of the network, especially in a large network setup.  Further, low complexity distributed relay selection protocols are desirable since computational complexity and the overhead of the optimal centralized scheduling are extremely high.

In recent years, distributed solutions for resource allocation problems have been proposed using opportunistic carrier sensing \cite{Zhao2005, chen2007integrated, Cohen2009, Cohen_2010_TSP, Yaffe2010, LeshemZehavi2012, Naparstek2013, Naparstek2014}.   In \cite{Zhou2008}, a power allocation is used to optimize either the energy consumption per bit or the lifetime of the network. The  distributed implementation of this protocol exploits the  timer-based relay selection  proportional to the instantaneous channel \cite{Bletsas2006}.  It should be noted that a relay scheduling scheme with power control requires a solution involving  linear programs at each node.  Moreover, the circuit power consumption is not considered in these papers,  so that the solution may be sub-optimal. It is noted that when the operating transmission range is relatively small, the power consumed in the baseband processor and in other components in the RF chain might be much greater than the power consumed by the transmitter power amplifier.    Moreover, minimizing transmission power  does not correspond directly to reduction in energy consumption due to  energy consumption in the circuitry of the device \cite{Buzzi2015_survey_JSAC}. 
 
 Device to device (D2D) communication has emerged as  a potential technique to reduce the energy consumption of mobile networks and is considered as one of the key  techniques for the LTE (Long Term Evolution) based cellular  networks \cite{Feng2013, Asadi2014, Tehrani2014_Comm_Mag,Mach2015_COMST}. The D2D communication enables devices to directly communicate with each other without transversing the data through a base station (BS).  Recent works  \cite{Song2015_CN, Deng2015GC,Chaaban2015, Hourani2016,Asadi2017}  show that the devices can work as relays to avoid  deployment and maintenance of conventional relays. Relay assisted D2D can  leverage maximum potential of the D2D communication in  various practical scenarios, such as when the distances between  D2D pairs are longer or when the devices experience strong  fading. In this context, an opportunistic scheduling of devices is shown to perform well for improved spectral efficiency \cite{Song2015_CN}. The authors in \cite{Hourani2016}  provide a geometrical zone for energy efficient D2D relaying. Furthermore, an experimental analysis of out-band D2D relaying  scheme is presented to integrate D2D communications in cellular network \cite{Asadi2017}. Performance for D2D communication under different fading environments is recently investigated (see \cite{Li2013, Chun2017}, and references therein).   In \cite{Messier2010}, the author investigated the performance of opportunistic techniques for average signal-to-noise ratio improvements in wireless sensor network under large scale channel effects. The log-normal shadowing environment is common in various practical scenarios including shopping malls, offices, university building \cite{Liu2015}.

In this paper, we  analyze the opportunistic relaying in the paradigm of D2D communication and study the opportunistic schemes   with a single relay selection to minimize energy consumption of the devices in a wireless network. The D2D communication ensures that the energy consumed for the data relaying is negligible, and that the selection of the relaying device exploits the diversity from neighboring devices in order to minimize the energy consumed by the devices for  forwarding the data to the BS. The opportunistic  device-select relaying (DSR) protocol  selects a single device based on an instantaneous transmission energy including the energy consumed in the circuitry of the devices and overhead energies required for relay selection.    The DSR is implemented in a distributed way using the opportunistic carrier sensing algorithm for the selection of the device. This builds on the single hop protocol developed in \cite{Cohen_2010_TSP}, with a proper adaptation to the two-hop case as well as an adaptation of the opportunistic scheduling considering the transmission energy.

 We derive closed form analytical bounds law on the expected  energy consumption of the DSR protocol where the devices experience log-normal shadowing as well as path loss inside a building or densely populated areas with respect to the BS whereas the devices close to the source device  experiences strong signal under Rayleigh fading channel. We also  derive a scaling law on the expected energy consumption with respect to the number of  devices, and show that the proposed scheme improves the energy efficiency of the network comparing to the direct communication using only few devices of the network.  A simple power control  at the relays is presented to minimize the transmission energy. Finally, we demonstrate the effectiveness of the protocols with a numerical analysis, using  parameters from realistic mobile networks, and we compare it to state of the art solutions.

The rest of this paper is organized as follows. Section II defines the system model. The proposed protocol for D2D relaying is described in Section III. Performance analysis of the DSR protocol is presented in Section IV.  Section V provides performance evaluation using common scenarios considered for cellular networks. Section VI concludes the paper.


\section{System Model}
\begin{figure}[t]
	\begin{center}
		\includegraphics[scale=0.7]{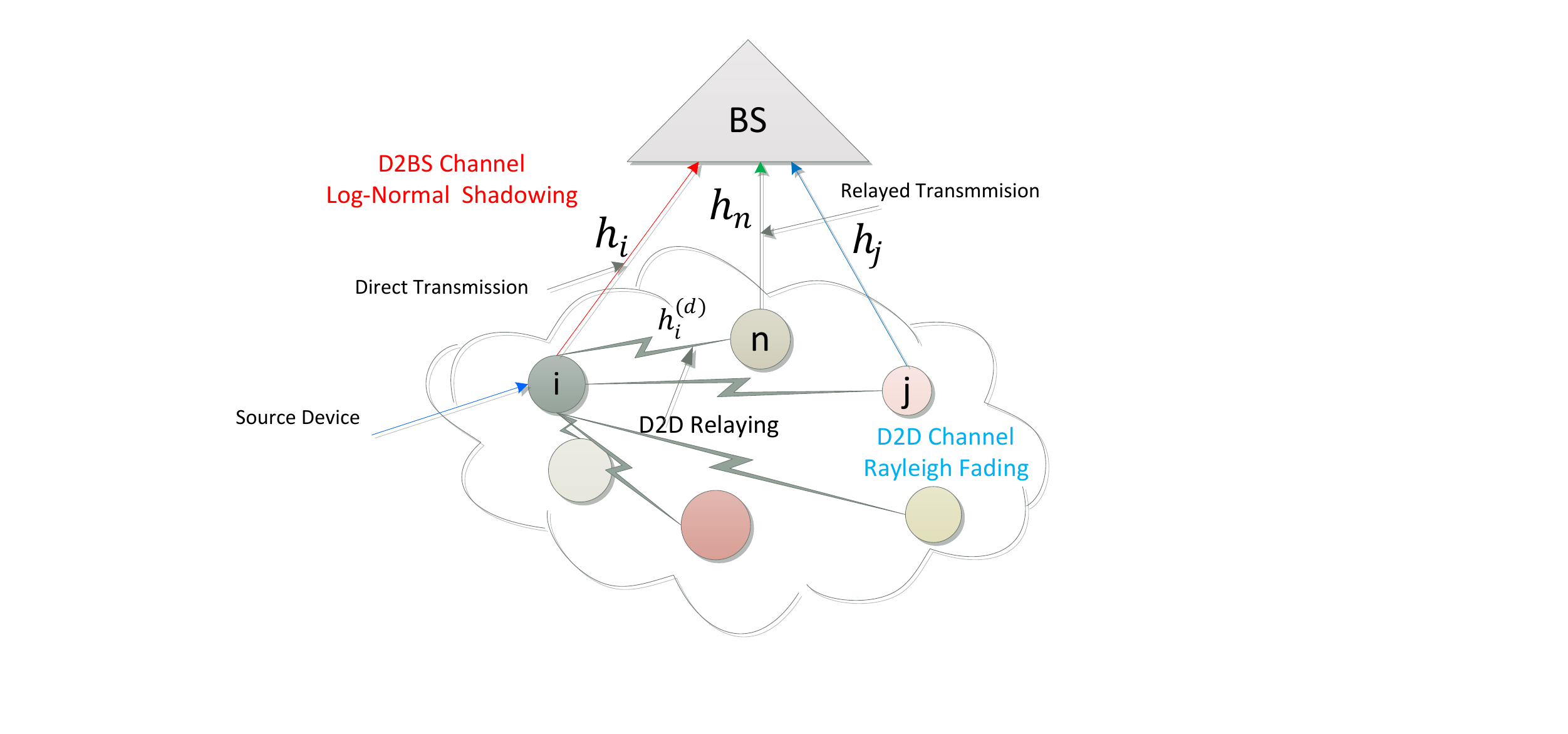}
	\end{center}
\vspace{-14mm}
	\caption{D2D relaying in the uplink communication of a single cell network.}
	\label{fig:model}
\end{figure}
We consider a single-cell network  in which a BS serves $N$ single-antenna devices, which  are uniformly distributed in the network.   We consider data transmission to the BS in the uplink direction.  We consider the wireless network where the devices experience log-normal shadowing as well as path loss inside a building or densely populated areas with respect to the BS. This type of fading environment is common in various practical scenarios including shopping malls, offices, university building, etc. These scenarios impose significant constrains and limitations on communication with the far away destination, because of the signal attenuation due to walls. This drawback becomes much more pronounced at high frequencies, such as millimeter-wave communications (a key enabler for next generation networks \cite{Qiao2015}), where quality of direct transmission is weak.

 We focus on a two-hop transmission model, where a  source device can either transmit data directly to the BS or relay the data to a nearby device, which forwards the data  to the BS, as depicted in Fig.~\ref{fig:model}. In a transmission slot, all devices are allocated with a  unique resource block (RB)  by the BS to avoid  inter-carrier interference between devices. 

In the direct transmission, the received signal at the BS from the $i$-th source device is given as:
\begin{align}
y^{\rm BS} = \sqrt{P}h_ix_i+w,
\label{eq:ybs}
\end{align}
where $P$ is the  transmit power, $h_i$ is the channel response between the $i$-th device and the BS, $x_i$ is the transmitted signal with unit power $\mathbb{E}[|x_i|^2]=1$,  and $w\sim {\cal{CN}}(0,{N_0})$ is  the zero-mean additive white Gaussian noise (AWGN) with power $N_0$. The additive noise contains  thermal noise and other interference terms (if any) which may increase the background noise level of the BS.

Since long term path loss dominates the short term fading, and  over longer time scales Rayleigh fading is averaged out, we model the  amplitude power $|h_i|^2$ as log-normal distributed:
\begin{align}
|h_i|^2 = GR_i^{-\alpha_{i}}\cdot 10^{\frac{S_i}{10}} ,~~ i= \{1,2,\cdots N\}
\label{eq:hn_sqr}
\end{align}
where $R_i$ is the distance from the $i$-th device to the BS, and $\alpha_i$ is the path loss coefficient. The term $S_i \sim {\cal{N}}(0,\sigma^2)$ is normal such that $10^{\frac{S_i}{10}}$ is log-normally  distributed. The parameter $\sigma$ is known as the dB spread or the shadowing factor. The term $G$ is the normalizing factor for the path loss.   Using (\ref{eq:hn_sqr}), we can represent the log-normal fading channel as normal taking the logarithm of channel gain as: $10\log_{10}|h_i|^2 = X_i\sim {\cal{N}}(10\log_{10} R_i^{-\alpha}+10\log_{10} G,\sigma^2)$.  

If the direct transmission  is not energy-efficient (e.g. due to shadowing effect reflected  in the channel $h_i$) the source device sends data to a relay device using D2D communication.  The received signal at the $n$-th relay  device is given as
\begin{align}
y_n^{(\rm d)} = \sqrt{P}h_{i}^{(\rm d)} x_i+v,
\label{eq:yj}
\end{align}
where $h_{i}^{(\rm d)}$ is the channel response  between the  $i$-th source device and the selected relay device,  and $v$ is AWGN with power $N_0$ including interference at the relay device.  Since the quality of signal received at the relay is high, a DF protocol is used at the relay to transmit the data from the source device to the BS.  It is noted that all  devices use different RBs separated in time and frequency, and thus there is no interference even if a single relay device receives signal from multiple source devices as these are sent at different RBs.

We assume  fading amplitude $|h_{i}^{(\rm d)}|$ between the $i$-th source device and the relay device as  Rayleigh distributed such that
\begin{align}
|h_{i}^{(\rm d)}|^2 = r_{i}^{-\alpha^{(\rm d)}}\cdot F_{i},
\end{align}
where   $F_{i}$ follows the exponential  distribution, $r_{i}$ is the distance from the $i$-th source device to the selected relay device, and $\alpha^{(\rm d)}$ is the path loss exponent between them. Since devices are close each other in D2D communication, the relay devices receive signal at a very high SNR, and thus consume negligible energy compared with the direct transmission.

\section{Description of the Protocol}
In this section, we  describe the proposed protocol and the criteria used in which devices cooperate with each other to minimize transmission energy and to improve  energy efficiency of the network. Although the description of the protocol can fit in many source-relay-destination frameworks, we focus on an uplink communication of the  wireless network. Without loss of generality, we assume that there is a single source device in every transmission slot. 

\subsection{Energy Consumption}
We assume transmissions of packets with  a fixed length of $L$ bits in a RB transmitted by the device to the BS in each transmission slot. We assume that the fading channel is constant for the duration of the RB of the source device. We assume that all devices transmit with equal power $P$. We denote the circuit power  by $P_i^{\rm ckt}$ for the $i$-th device. Since  the power dissipated in  the transmitter and receiver circuits is different for different devices,  we assume that the circuit power transmission of the devices is uniformly distributed between  $P_{\rm min}^{\rm ckt}$  and  $P_{\rm max}^{\rm ckt}$.  We  denote the initial  energy  of the $i$-th device by  $E_{i}^{\rm in}$.

Using \eqref{eq:ybs}, the energy consumed by the $i$-th source device to transmit its data directly to the BS  is:
\begin{align}
\begin{split}
E_{i} = (P+P^{\rm ckt}_i)\cdot \frac{L}{B_{}\log_2(1+\gamma_i)}
=  \frac{\eta_1}{10\log_{10}(1+\gamma_i)} + \frac{\eta_2P^{\rm ckt}_i}{10\log_{10}(1+\gamma_i)}
\end{split}
\label{eq:en}
\end{align}
where $B$ is the transmission channel bandwidth, $\eta_1=10\log_{10}(2)PL/B$, $\eta_2^{\rm}=\eta_1/P$, and $\gamma_i=\frac{|h_{i}|^2P_{}}{N_0}$is the received signal-to-noise-ratio (SNR) at the BS when the signal is transmitted from the $i$-th device.   Note, that as we assume the DF relaying,  the energy is computed based on optimal coding scheme between the source and the relay, and a different rate is used for the second hop.

From \eqref{eq:en}, it can be seen that  a minimization of the  transmit power does not correspond to the minimum energy transmission. Here, we provide a simple optimal transmit power control strategy using local measurements.  Using first order derivative of (\ref{eq:en}) with respect to $P$, an  optimal transmit power $P^*$ can be obtained using the solution of the following logarithmic equation:
\begin{align}
(1+\frac{|h_{i}|^2}{N_0}P^*) \log(1+\frac{|h_{i}|^2}{N_0} P^{*})=\frac{|h_{i}|^2}{N_0}(P^{*}+P_i^{\rm ckt})
\label{eq:opt_power}
\end{align}
This optimal transmit power minimizes the transmission energy in the presence of circuit power.

Using \eqref{eq:yj},  the energy consumed by the D2D communication to relay a data of $L$ bits is:
\begin{align}
\begin{split}
E_{i}^{(\rm d)} = 
\frac{\eta_1^{(\rm d)}}{\log_{e}(1+\gamma_i^{(\rm d)})}+ \frac{\eta_2^{(\rm d)}P^{\rm ckt}_i}{\log_{e}(1+\gamma_i^{(\rm d)})}
\end{split}
\label{eq:ed}
\end{align}
where $\eta_1^{(\rm d)}=\log_{e}(2)P^{\rm d}L/B$, $\eta_2^{(\rm d)}=\eta_1^{(\rm d)}/P^{(\rm d)}$, and $\gamma_i^{(\rm d)}=\frac{|h_{i}^{(\rm d)}|^2P^{(\rm d)}}{N_0^{(\rm d)}}$ is the SNR at the relay device when the signal is transmitted from the $i$-th source device. Similarly, an optimal transmit power can be obtained for the relaying phase in the  D2D communication.

\subsection{Device Selection Metric}

For the relay selection, we consider  two main components of the energy consumption:  energy $E_i^{(\rm d)}$ computed in \eqref{eq:ed} for the D2D communication  and energy  $E_i^{}$ computed in \eqref{eq:en} to transmit the data to the BS from the $i$-th relay device. 

In contrast to the conventional relay selection schemes based on the SNR or fading channels \cite{Bletsas2006}, \cite{delima2017}, the relay selection criteria for the proposed protocol is based on the total  consumed energy as:
 \begin{align}
n=\argmin_{1\leq i\leq N}  \{E_i+ E_i^{(\rm d)}+E_{\rm ov}^{\rm RELAY}\}, 
\label{eq:select1}
\end{align}
 where  $E_{\rm ov}^{\rm RELAY}$ is the overhead energy required for relay selection in the case of D2D communication. The dual-hop DSR selection criteria in \eqref{eq:select1} is useful when both hops are symmetric and consumes a similar amount of energy.  However, the scenario considered in this paper is highly imbalanced in both the hops because the  energy consumption of the D2D relaying (strong signal due to proximity of the devices) is lower than the energy consumed in the second hop (users are generally at a larger distance from the BS). Hence, the selection criteria in \eqref{eq:select1} is reduced to relay selection based on  the energy consumption of the second hop only: 
 \begin{align}
n=\argmin_{1\leq i\leq N}  \{E_i\}.  
\label{eq:select2}
\end{align}
We define the single-hop  selection criteria in \eqref{eq:select2} as the DSR.  Since we use D2D communication in the first hop, the single hop relay selection in \eqref{eq:select2} is near-optimal and  depends only  on the channel quality between devices and the BS (which is available for the existing operation of the mobile network). This avoids energy overhead to estimate the channel using the  request-to-send (RTS) signal  in the first hop and clear-to-send (CTS) signal in the second hop as required in \cite{Zhou2008}.

In what follows, we describe a low complexity implementation of   DSR  protocol since a centralized implementation of the  protocols  would be  consuming  large energy overhead  due to control signaling. 
\begin{figure}[t]
		\begin{center}
	\subfigure[Timing diagram and resource block allocation.]{\includegraphics[scale=0.60]{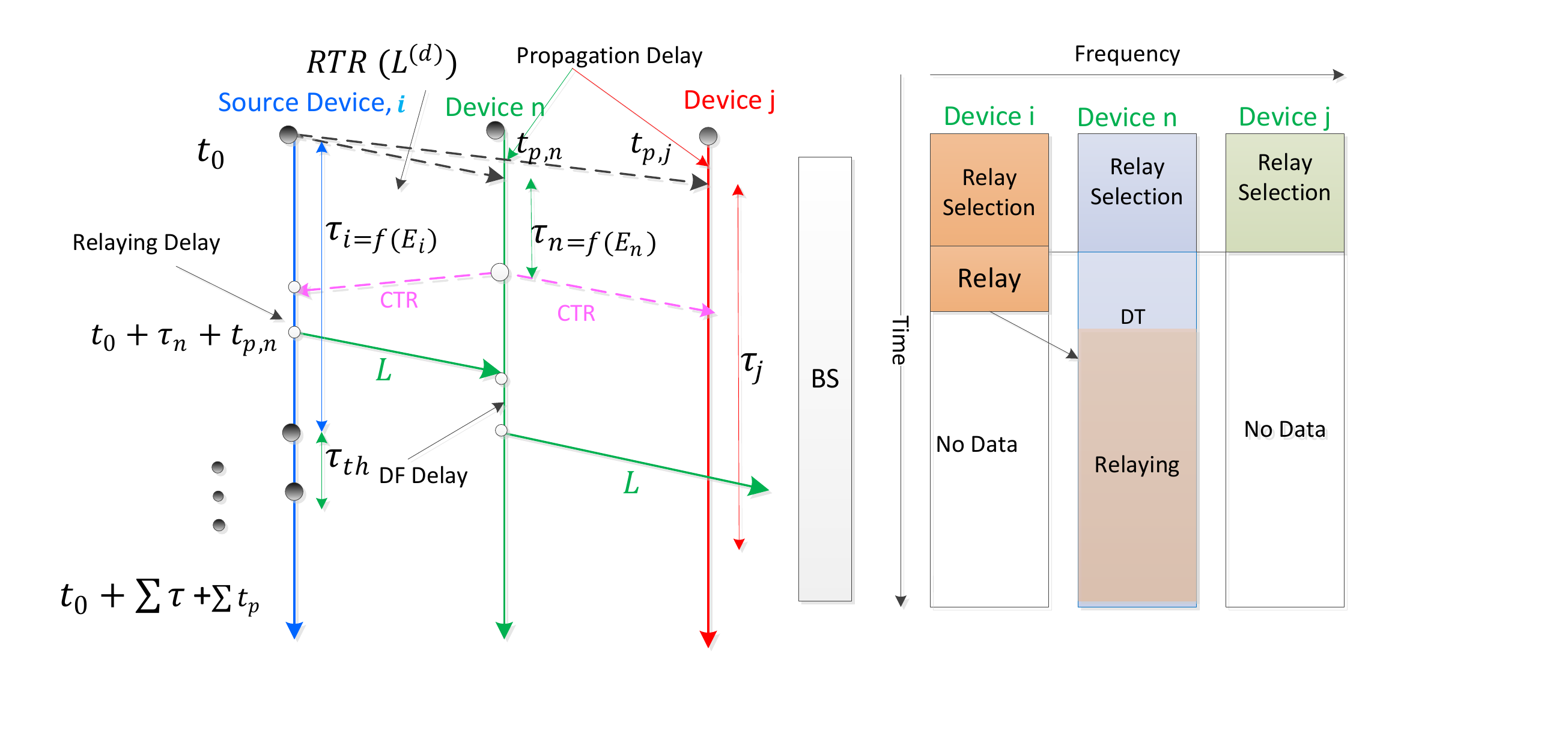}}
		\subfigure[Function $f(E)$.]{\includegraphics[scale=0.65]{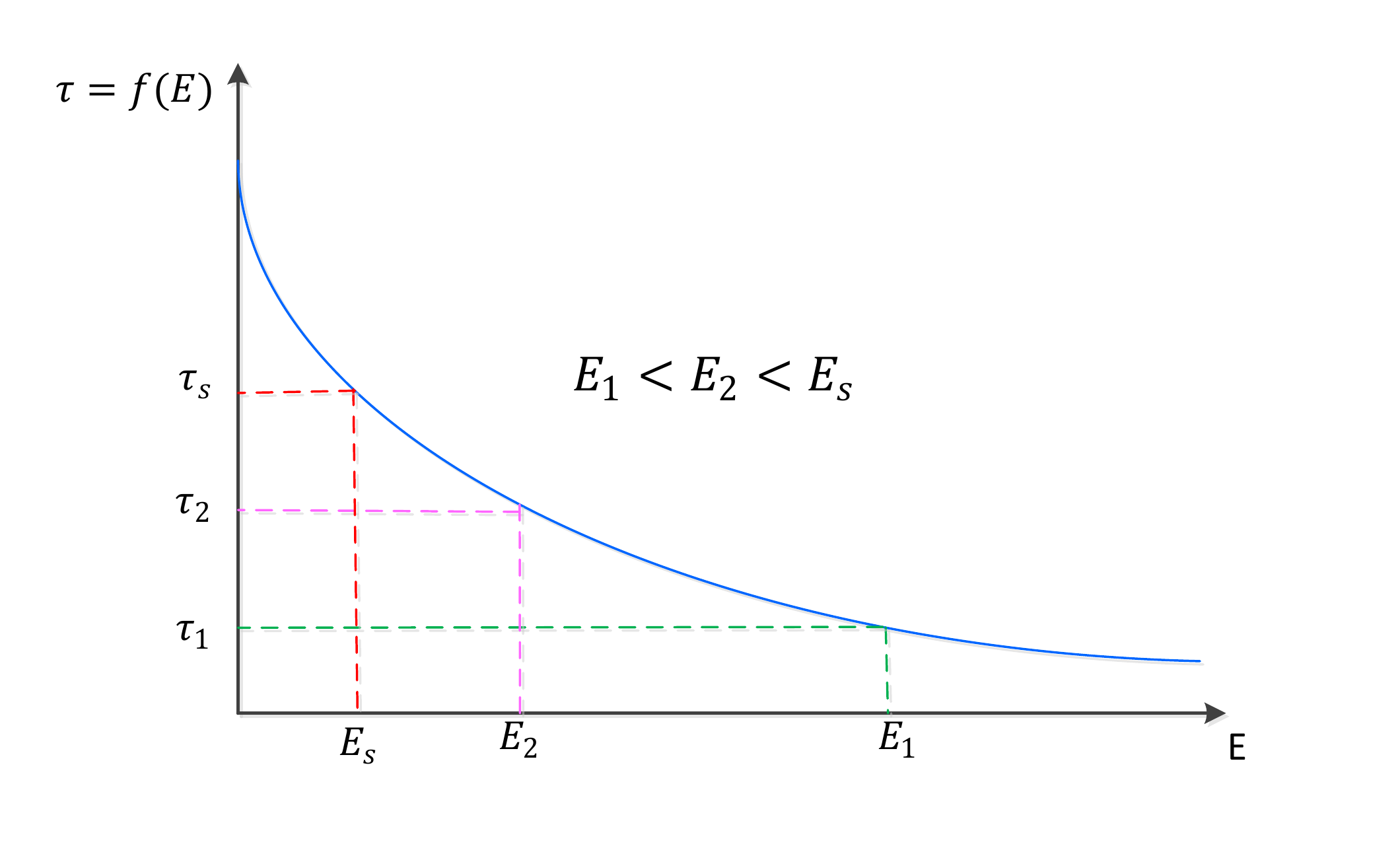}}
	\caption{Schematic diagram of the DSR  protocol for three devices with transmission energy $E_n<E_j<E_i$.}
	\label{fig:scheme2}
	\end{center}
\end{figure}
\subsection{Distributed Protocol}
For distributed implementation of the proposed protocols, we use the back-off principle of the carrier sensing multiple access (CSMA) in the TCP/IP MAC layer supported with the transmission energy  from the PHY. We define an increasing function $f(E)$ designed judiciously (see Fig. 2a) such that  back-off time $\tau_i=f(E_i), i=1 \cdots N$ of the devices computed with distinct energy index $E_i, i=1 \cdots N$  (given in equation \eqref{eq:en})  provides enough back-off range to minimize collisions without incurring high delay by the use of $\tau$. In the seminal paper, Blestsas et. al \cite{Bletsas2006} describe a timer-based distributed protocol for relay selection (controlled by the BS with RTS and CTS signals) using instantaneous channel information of both hops. Zhou et. al \cite{Zhou2008} use the protocol in \cite{Bletsas2006} for the relay selection using power control at each relays for energy-efficient transmissions. Our protocol differs from \cite{Bletsas2006,Zhou2008}  as follows: (i) our protocol directly minimizes energy at each transmission instead of diversity gain for rate performance using the instantaneous channel \cite{Bletsas2006} or power control \cite{Zhou2008}; (ii) no control signaling from the BS for relay selection is needed; (iii) our proposed protocol reduces hidden node problem due to the D2D communication.
The developed DSR protocol is implemented in following steps (see Fig.~\ref{fig:scheme2}):

\subsubsection{\bf Request to Relaying (RTR)}
First, the $i$-th source device sets its back-off time to $\tau_i=f(E_i)$ and broadcasts an  RTR message (with fields such as user ID) with a transmit power to be received by the devices in close proximity.     The transmit power for the RTR message can be controlled to minimize the overhead energy and avoid selecting relays incurring high energy consumption in relaying. It is noted that structure of RTR is quite different from RTS which contains many control bits.  All the devices are capable to decode the RTR message with the CSI estimated  using the RTR message if not known. The CSE is known if devices are already in the discovery mode  complaint with the proximity services of  3GPP-LTE \cite{Lin2014}. The RTR transmission costs an energy consumption $E_{\rm tx}^{\rm RTR}$ to the source device. The energy overhead in decoding the RTR per device is  $E_{\rm rx}^{\rm RTR}$.

The source device waits for a reply from a potential relay for a duration of $\tau_i+\tau_{\rm th}$,  where $\tau_{\rm th}$ is an additional delay to compensate for the propagation delays in D2D communication. This delay corresponds to relay selection overhead, as depicted in Fig.~ \ref{fig:scheme2}. If the device does not  receive a reply from any device for relaying in the time limit of $\tau_i+\tau_{\rm th}$,  it directly  transmits to the  BS (step 4),  otherwise the data is transmitted through a relay.

\subsubsection{\bf Distributed Relay Selection}
Upon the receipt of a RTR message from the source, each device sets its back-off time to $\tau_j=f(E_j), j\cdots N-1$. In the opportunistic DSR scheme, the $n$-th  device selected using the criteria in \eqref{eq:select2} has the lowest back-off time, and hence occupies the channel first by responding  to the source with a CTR message after a waiting period $\tau_n< \tau_j,  n\neq j$.   Once the selected device transmits the CTR message to the source, all other devices overhear the CTR message (or just a busy tone), and  quit the process of relay selection for the given request from the $i$-th source device.   The overhead energies for a response from relay device are:  transmission of  CTR message  $E_{\rm tx}^{\rm CTR}$ and  reception of   CTR message  $E_{\rm tx}^{\rm CTR}$.

 \subsubsection{\bf Source to Relay Transmission}
 Upon the successful decoding of the CTR message,  the source device  sends the data packet $L$ to the selected relay device with a transmit energy cost $E_{i}^{\rm d}$  as computed in \eqref{eq:ed}.
 Using the DF protocol, the selected relay device decodes the data from the source device, encodes it, and transmits to the BS. The DF protocol requires the CSI at the relay device. This can be estimated using the RTR message from the source device after the decision on relay selection.   The energy overhead at this stage is: CSI estimation energy  $E^{\rm CSI}$,  transmit energy cost $E_{i}^{\rm d}$,  decoding energy $E^{\rm DEC}$, and  encoding energy $E^{\rm ENC}$. 
 \subsubsection{\bf Data Transmission}
Finally, transmission of data is accomplished by direct transmission from the source or by the relay device. The energy consumption in this phase is $E_{i}$ as computed in \eqref{eq:en}.
 \subsection{Challenges of Distributed Implementation}
Although distributed implementation simplifies the relay selection and transmission comparing to the centralized method, it poses few challenges that needs consideration. 
We now describe some practical challenges, which may impact the distributed implementation, and their solutions as follows:
 \subsubsection{Coexistence of direct and relay transmission}
All devices have unique RB assigned by the eNB and relay devices use the RB of the source device for both D2D communication and data transmission. If a single device happens to act as the source for its own data and as the relay for other source, the data transmission can be done simultaneously.

 \subsubsection{Multiple users with same back-off time} As described in \cite{Bletsas2006}, the probability that two users have equal back-off time is zero. However, there might be scenarios where difference between the multiple back-off time can be  lower causing collision of CTR messages at the source device.  However due to different propagation delays  probability of this case is very low. In any case, the source device directly transmits after waiting for a specified time.

 \subsubsection{Hidden node problem}The broadcast of RTR message informs devices in proximity about relay selection. However, some devices overhearing RTR can be hidden from other relay devices.  Hence, a hidden device from already selected relay device may still be in the contention for the channel and sends CTR to the source device after its back-off time. This is avoided by the transmission of data from the source device which  is overheard by all devices in proximity.

 \subsubsection{Transmission delay} The relay selection based on CSMA increases an additional delay corresponding to the back-off time $\tau $. 	
In order to limit the additional delay in the relay selection process or delay due to unsuccessful reception of CTR (in case two devices have almost the same back-off time $\tau$), the source device transmits its data directly to the eNB after waiting  $\tau_i+\tau_{\rm th}$, where  $\tau_i$ is the back-off time of the source device and $\tau_{\rm th}$ is an additional delay to wait for successful reception of a CTR message. However, a use of the relay with best channel significantly  reduces time to transmit the data to the BS (higher throughput) and thus leads to a reduced transmission time, compared to the direct transmission.

In the following sections, we analyze the performance of the  proposed scheme by deriving bounds on the expected energy consumption, and by means of simulation results in a scenario of a realistic mobile network.

\section{Performance Bounds of Opportunistic DSR }

In this section, we discuss various components of the consumed energy, derive bounds on expected energy consumption of direct transmission and relay transmission, and perform an analysis on overhead energy of D2D relaying.
\subsection{Energy Consumption of  DSR}
Given the steps of distributed relaying described in the previous section, the total energy consumption using the DSR   protocol is:
\begin{align}
\begin{split}
E^{\rm DSR}=p(E^{\rm RELAY}+ E^{\rm D2D}+ E_{\rm ov}^{\rm RELAY})+(1-p)(E^{\rm DT}+ E_{\rm ov}^{\rm DT})
\end{split}
\label{eq:dsr_protocol}
\end{align}
where $p$ is the probability of the relay-assisted data transmission, $E^{\rm RELAY}$ is the energy consumed by the selected relay to transmit the data packet to the BS, $E^{\rm D2D}$ is the data transmission energy by the source device to another relay in D2D communication,   $E_{\rm ov}^{\rm RELAY}= E_{\rm tx}^{\rm RTR}+(N-1)E_{\rm rx}^{\rm RTR}+E_{\rm tx}^{\rm CTR}+E_{\rm rx}^{\rm CTR}+E^{\rm CSI} +E^{\rm DEC}+E^{\rm ENC}$ is the overhead energy required for relay selection in the case of D2D communication, $E^{\rm DT}$ denotes the energy consumed  for data transmission directly to the BS when the direct transmission is found to be more energy-efficient  than the relay-assisted transmission, and  $E_{\rm ov}^{\rm DT}=E_{\rm tx}^{\rm RTR}+(N-1)E_{\rm rx}^{\rm RTR} $ is  overhead energy for the relay selection. It is noted that the direct transmission without DSR protocol does not incur any overhead energies. However, the overhead energies in  the DSR protocol are low since the signaling involved is very short and the signaling messages are sent to other local devices with very low power or at a very high data rate.

We compare the transmission energy of the DSR protocol $E^{\rm DSR}$ with the component of  the transmission energy for only the direct transmission $E^{\rm DT}$.  We analyze the  components of energy consumption in \eqref{eq:dsr_protocol}, and  demonstrate  that the DSR protocol consumes less energy than the direct transmission.

\subsection{Expected  Energy Consumption of  Direct Transmission}
We derive an expression on the expected consumed energy without D2D relaying. In this scenario, each device transmits its own data to the BS.  Using a simple inequality,  $10\log_{10}(\gamma)\le 10\log_{10}(1+\gamma)\le 1+10\log_{10}(\gamma)$ in  (\ref{eq:en}), we get bounds on energy  consumption  for  the direct transmission as
\begin{align}
\frac{\eta_1+\eta_2P^{\rm ckt}}{1+X} \le E^{\rm DT} \le  \frac{\eta_1+\eta_2P^{\rm ckt}}{X},
\label{eq:E_0}
\end{align}
where $X=10\log_{10}(\gamma)$. Since $\gamma$ is log-normal distributed with a spreading parameter $\sigma^2$ in dB, $X\sim{\cal{N}}(\bar{\gamma},\sigma^2)$ with $\bar{\gamma}= 10\log_{10} R^{-\alpha}+10\log_{10} G+10\log_{10}P/N_0$ is the average SNR per device in dB. Considering different types of user devices in a network, we model the circuit power to be uniformly distributed between $P_{\rm min}^{\rm ckt}$ and $P_{\rm max}^{\rm ckt}$ representing minimum and maximum circuit transmit powers, respectively.  We  assume  circuit transmit power  independent of  transmit power since   the variation on the circuit transmit power with the transmit power is insignificant.   We compute the upper bound, but similar expression hold for the lower bound, replacing $\bar{\gamma}$ with $\bar{\gamma}+1$. Taking expectation in \eqref{eq:E_0} and noting the independence between numerator and denominator terms,  we get an upper bound on the expected energy consumption with direct transmission defined as
\begin{align}
\begin{split}
\bar{ E}^{\rm DT}\leq \mathbb{E}[ \eta_1+\eta_2P^{\rm ckt}]\mathbb{E}[\frac{1}{X}]
  =\left(\eta_1+\eta_2\mathbb{E}[P^{\rm ckt}]\right)\frac{1}{\sqrt{2\pi}\sigma}\int_{\gamma_{\rm th}}^{\infty}\frac{1}{x } e^{-\frac{(x-\bar{\gamma})^2}{2\sigma^2}} \mathrm{d}{x} 
 \end{split}
\label{eq:E_direct1}
\end{align}
where $\gamma_{\rm th}$ in dB is a SNR threshold   above which the communication occurs. The threshold SNR is selected to achieve a minimum data rate requirement below which communication is possible.

\begin{my_theorem}
	\label{theorem:dt}
 If  $P_{\rm min}^{\rm ckt}$ and $P_{\rm max}^{\rm ckt}$ are minimum and maximum circuit transmit    power of all devices, respectively, $\gamma_{\rm th}$ is the threshold SNR in dB, and $\eta_1=10\log_{10}(2)PL/B$, $\eta_2^{\rm}=\eta_1/P$, then bounds on the expected energy with direct transmission in a log-normal fading channel with  average SNR  $\bar{\gamma}$ and variation $\sigma$ in dB are given as
	\begin{align}
	\begin{split}
&\frac{(\eta_1+0.5\eta_2(P_{\rm max}^{\rm ckt}+P_{\rm min}^{\rm ckt}))} {(\bar{\gamma}+1)} \exp{(\frac{\sigma^2}{2(\bar{\gamma}+1)^2})}  Q(\frac{{\sigma}}{(\bar{\gamma}+1)}+\frac{(\gamma_{\rm th}-\bar{\gamma}-1)}{\sigma})   \leq\bar{E}^{\rm DT}\\&\leq(\eta_1+0.5\eta_2(P_{\rm max}^{\rm ckt}+P_{\rm min}^{\rm ckt}))[{\cal{I}}_1^{\rm DT}(\bar{\gamma},\sigma)+{\cal{I}}_2^{\rm DT}(\bar{\gamma},\sigma)], 
	\end{split}
	\label{eq:theorem:dt}
	\end{align}
where
	\begin{align}
\begin{split}
&{\cal{I}}_1^{\rm DT}(\sigma)
=\frac{\sigma}{\sqrt{2\pi}({2\sigma^2+\bar{\gamma}^2})}\Big[2\sqrt{2}\sigma\log_e(\frac{\bar{\gamma}}{\gamma_{\rm th}})\log_e(1+(\frac{\bar{\gamma}-\gamma_{\rm th}}{\sqrt{2}\sigma})^2)+\arctan{(\frac{\bar{\gamma}-\gamma_{\rm th}}{\sqrt{2}\sigma})^2}\Big]\\&
{\cal{I}}_2^{\rm DT}(\sigma)=\frac{\exp[-\bar{\gamma}^2/2\sigma^2]}{4\sqrt{2\pi}\sigma}[2\pi {\rm Erfi}(\frac{\bar{\gamma}}{\sqrt{2}\sigma})-2E_1(\frac{\bar{\gamma}^2}{2\sigma^2})+\log_e(\frac{\bar{\gamma}^2}{2\sigma^2})+4\log_e(\frac{\sqrt{2}\sigma}{\bar{\gamma}})-\log_e(\frac{\sigma^2}{\bar{\gamma}})]	\\&
\end{split}
\label{lemma1_direct_ub}
\end{align}

	\end{my_theorem}

\begin{IEEEproof}
The proof is presented in Appendix A.
	\end{IEEEproof}
	It can be seen from the bounds in \eqref{eq:theorem:dt}  that a higher average SNR reduces the energy consumption of the direct transmission. The  derived expressions are computable in terms of known functions.

Now, we analyze the relaying protocol where a best relay is selected to minimize consumed energy, and show that the proposed scheme consumes less energy than the direct transmission  and the energy gain increases with the number of participating relay devices.
 \subsection{Expected Energy Consumption for Relay Selection}
 To simplify  the model, we assume that the relaying devices are in the vicinity of the source, so that the path loss of all possible relays are similar, but  spread enough to experience independent shadowing. To compute energy consumed by the relaying, we derive expression for the expected energy consumed $E^{\rm RELAY}$  by the device to the BS  in log-normal fading with the selection criteria defined in \eqref{eq:select2}.  Using the selection criteria in (\ref{eq:select2}) for the log-normal shadowing (device to eNB transmission) in (\ref{eq:E_0}), we get:
 \begin{align}
 \begin{split}
 E^{\rm RELAY}\le\frac{\eta_1+\eta_2P_{(1)}^{\rm ckt}}{X_{(n)}}
 \label{eq:En1}
 \end{split}
 \end{align}
 where $X_{(n)}=\max(X_1,X_2,X_3,\cdots,X_N)$, $X_i=10\log_{10}(\gamma_i)$, and $P_{(1)}^{\rm ckt} =\min(P_{1}^{\rm ckt},P_{2}^{\rm ckt},P_{3}^{\rm ckt},\cdots,P_{N}^{\rm ckt})$ 
Since we assume similar path loss for all device (i.e., relays are in the vicinity of the source) the $X_i$ are i.i.d.; it follows from standard order statistics that the CDF of $X_{(n)}$ is given as  $F_{X_{(n)}}(x) = [F_X(x)]^N$, where $F_{X}(x)= [1/2+1/2\rm {erf}(\frac{x-\bar{\gamma}}{\sqrt{2\sigma^2}})] $ is the CDF of normal distribution. The PDF of the maximum $X_{(n)}$ is $f_{X_{(n)}}(x) =N  [F_X(x)]^{N-1} [f_{X}(x)]$
where $f_{X}(x) =\frac{1}{\sqrt{2\pi\sigma^2}}e^{-\frac{(x-\bar{\gamma})^2}{2\sigma^2}}$ is the PDF of normal distribution.
The expected consumed energy with the best select relay scheme $\bar{E}^{\rm RELAY} =\mathbb{E}[E^{\rm RELAY}]$ is expressed as:
\begin{align}
\begin{split}
\bar{E}^{\rm RELAY} \le \mathbb{E}[ \eta_1+\eta_2P_{(1)}^{\rm ckt}]  \int_{\gamma_{\rm th}}^{\infty}\frac{ N}{x}[F_X(x)]^{N-1} [f_{X}(x)]\mathrm {d} x,
\label{eq:dsr_main1}
\end{split}
\end{align}

Using integration in parts  and  $F_X(x) = Q(\frac{\bar{\gamma}-\gamma_{\rm th}}{\sigma})$, we can represent  (\ref{eq:dsr_main1}) as:
\begin{align}
\begin{split}
\bar{E}^{\rm RELAY} \le\mathbb{E}[ \eta_1+\eta_2P_{(1)}^{\rm ckt}] \Big({\cal{I}}_{1}^{\rm RELAY}(N,\sigma)+{\cal{I}}_{2}^{\rm RELAY}(N,\sigma)-\frac{1}{\gamma_{\rm th}} Q^N(\frac{\bar{\gamma}-\gamma_{\rm th}}{\sigma})\Big )
\end{split}
\label{eq:dsr_main2}
\end{align}
where
\begin{align}
\begin{split}
& {\cal{I}}_{1}^{\rm RELAY} (N,\sigma) =\int_{\frac{\gamma_{\rm th-\mu}}{\sigma}}^{0}\frac{1}{(x\sigma+\bar{\gamma})^2}(1-Q(x))^N\mathrm {d} x \\&
{\cal{I}}_{2}^{\rm RELAY} (N,\sigma) =\int_{0}^{\infty}\frac{1}{(x\sigma+\bar{\gamma})^2}(1-Q(x))^N\mathrm {d} x 
\end{split}
\label{eq:dsr_main3}
\end{align}
Computable bounds and analytical expressions can be obtained  from (\ref{eq:dsr_main2})
using Chernoff bounds (upper and lower as given in \cite{Chang2011_lower}) on the Q-function that makes integration tractable and in a closed form.

\begin{my_theorem}
	\label{theorem:relay}
 If  $P_{\rm min}^{\rm ckt}$ and $P_{\rm max}^{\rm ckt}$ are minimum and maximum circuit transmit   power of all devices, respectively, $\gamma_{\rm th}$ is the threshold SNR in dB, and $\eta_1=10\log_{10}(2)PL/B$, $\eta_2^{\rm}=\eta_1/P$, then the expected energy consumption with  a single relay selection from $N$ devices in a log-normal fading channel  with  average SNR  $\bar{\gamma}$ and variation $\sigma$ in dB is given as
	\begin{align}
		\begin{split}
			\bar{E}^{\rm RELAY} \le (\eta_1+\frac{\eta_2(P_{\rm max}^{\rm ckt}+NP_{\rm min}^{\rm ckt})}{	N+1}) \Big({\cal{I}}_{1}^{\rm RELAY}(N,\sigma)+ {\cal{I}}_{2}{\rm ^{RELAY}(N,\sigma)}- \frac{1}{\gamma_{\rm th}} Q^N(\frac{\bar{\gamma}-\gamma_{\rm th}}{\sigma})\Big)
		\end{split}
		\label{eq:theorem:relay}
	\end{align}
	where $Q$ denotes the standard Gaussian $Q$ function, and  
		\begin{align}
		\begin{split}
		{\cal{I}}_{1}^{\rm RELAY}(N,\sigma)\leq&
		\frac{\sigma}{(2)^N(2\sigma^2+N\bar{\gamma}^2)^2}\Bigg[{2\sigma^2(2\sigma^2+	N\bar{\gamma}^2)}\big(\frac{1}{\gamma_{\rm th}}-\frac{1}{\bar{\gamma}}\big) +{4N\sigma^2\bar{\gamma}}\log\big(\frac{\gamma_{\rm th}}{\bar{\gamma}}\big)+\\&{2N\sigma\mu}\log\Big(1+\frac{N}{2}(\frac{\bar{\gamma}-\gamma_{\rm th}}{\sigma})\Big)+{\sqrt{2N}(N\bar{\gamma}^2-2\sigma^2)}\arctan\Big({\sqrt{N/2}(\frac{\bar{\gamma}-\gamma_{\rm th}}{\sigma})\Big)}\Bigg]
		\end{split}
		\label{lemma2_I1_dsr_ub}
		\end{align}
		\begin{align}
		\begin{split}
		{\cal{I}}_{2}^{\rm RELAY}(N,\sigma)\leq& \sigma  \sum_{r=0}^{N}{{{N}/{2}}\choose{2r}}\frac{1}{4^{r}}\Psi(r,\sigma,\bar{\gamma})
		-\sum_{r=0}^{N}{{N/2}\choose{2r+1}}[f(\kappa)]^{2r+1}\Psi((2r+1)\kappa,\sigma,\bar{\gamma}), \\&
		\text{where}~~  f(\kappa)\leq \sqrt{e/2\pi}(\sqrt{\kappa-1})/\kappa, ~\text{and} ~ \Psi(r,\sigma,\bar{\gamma}) ~\text{is given in}~	\eqref{eq:prep_integral} ~\text{Appendix B}.
		\end{split}
		\label{lemma2_I2_dsr_ub}
		\end{align}

\end{my_theorem}

\begin{IEEEproof}
	The proof is presented in Appendix 	B.
	\end{IEEEproof}
The numerical analysis shows that	$I_{2}^{\rm RELAY}(N,\sigma)$ is dominant and bounds on the Q-function are not always accurate.  We improve the accuracy of analysis in Theorem 2 by using an approximation on  the Q-function in the following corollary.
\begin{my_corr}
	\label{corollary:I2}
Approximating the Q-function, a closed form expression on $I_{2}^{\rm RELAY}(N,\sigma)$ is:
\small
	\begin{align}
\begin{split}
&I_{2}^{\rm RELAY}(N,\sigma)\approx \sigma\sum_{k=0}^{N}{N\choose{k}}(-1)^k \Bigg(\frac{A(k)\gamma_{\rm max}}{\bar{\gamma}^2+\bar{\gamma}\sigma\gamma_{\rm max}}+\frac{B(k)}{\bar{\gamma}}\log(1+\frac{\sigma\gamma_{\rm max}}{\bar{\gamma}})+ C(k) \log|1+\frac{\gamma_{\rm max}}{\alpha(k)}|+D(k) \log|1+\frac{\gamma_{\rm max}}{\beta(k)}|\Bigg)\\&
A(k) = \frac{\sigma^2}{(\bar{\gamma}-\alpha(k)\sigma)(\bar{\gamma}-\beta(k)\sigma)},  B(k)=\frac{\sigma^2 (\alpha(k) \sigma+\beta(k) \sigma-2 \bar{\gamma})}{(\bar{\gamma}-\alpha(k) \sigma)^2 (\bar{\gamma}-\beta(k) \sigma)^2}, C(k) = \frac{1}{(\alpha(k)-\beta(k)) (\alpha(k) \sigma-\bar{\gamma})^2},\\& D(k) = \frac{1}{(\alpha(k)-\beta(k)) (\beta(k) \sigma-\bar{\gamma})^2},  \{\alpha(k), \beta(k)\}  =\big(-kq_2\pm\sqrt{k^2q_2^2-4kq_1q_2-4kq_1}\big)/2kq_1\\&
q_1=-0.4920, q_2= -0.2287,  q_3= -1.1893.
\label{lemma2_I2_dsr_appr}
\end{split}	
\end{align}
\end{my_corr}	
\normalsize
\begin{IEEEproof}
The	proof is presented in Appendix 	C.
\end{IEEEproof}
Finally, we derive an upper bound on the energy consumption and provide a scaling law for the reduction of energy with the number of devices.
\begin{my_theorem}
	\label{theorem:scaling}
If  $P_{\rm min}^{\rm ckt}$ and $P_{\rm max}^{\rm ckt}$ are minimum and maximum circuit transmit   power of all devices, respectively, $\gamma_{\rm th}$ is the threshold SNR in dB, and $\eta_1=10\log_{10}(2)PL/B$, $\eta_2^{\rm}=\eta_1/P$, then the expected energy with  a single relay selection from $N$ devices  in a log-normal shadow fading channel with  average SNR  $\bar{\gamma}$ and variation $\sigma$ in dB is upper bounded as:
\small
		\begin{align}
	\begin{split}
	\bar{E}^{\rm RELAY} \le \Big(\eta_1+\frac{\eta_2(P_{\rm max}^{\rm ckt}+NP_{\rm min}^{\rm ckt})}{	N+1}\Big) \Big( \frac{1}{2^N} \frac{1}{\gamma_{\rm th}} + \frac{1}{\sigma}\Big(\frac{1}{\bar{\gamma}+ \sigma\sqrt{c_M\ln(N)}}+\sum_{m=1}^{M-1}(\frac{1}{1+\kappa N^{(1-c_m)}}) (\frac{1}{\bar{\gamma}+\sigma \sqrt{c_{m-1}\ln(N)}}\Big)\Big)
	\end{split}
	\label{eq:theorem:scaling}
	\end{align}
	\normalsize
 where  $M$ is a positive integer, $\kappa=0.3885$ is a constant, and $ 0\leq c_m\leq1$, $c_0=0$, $m=1,2,\cdots M$. 
  Further, energy consumption scales as  
  \begin{align}
  \label{eq:scaling_law}
  \bar{E}^{\rm RELAY} = {\cal{O}}\big({\frac{\eta_1+\eta_2P_{\rm min}^{\rm ckt}}{\bar{\gamma}+\sigma\sqrt{c_M\ln(N)}}}\big).
\end{align}
\end{my_theorem}
\begin{IEEEproof}
The proof is presented in Appendix 	D.
\end{IEEEproof}
	From the scaling law in \eqref{eq:scaling_law}, it can be seen that the number of devices improves the energy consumption.
\subsection{Expected Energy Consumption for D2D Relaying}
As discussed in the protocol description (Section III), the  signaling involved in the relay selection is very short and incurs negligible energy overhead. This is illustrated through simulations in realistic scenarios in the next section. However, the energy overhead due to the data forwarding to the selected relay device in a fading environment and far apart from the source devices may increase the energy consumption significantly. This energy overhead can be minimized if  the selection of relay  is confined  only to the devices nearby to the source devices. In what follows, we analyze this energy consumption under Rayleigh fading channel, and show that the energy consumed  by relaying  is also negligible in the D2D communication range.

Under the Rayleigh fading channel model, the SNR $\gamma^{(d)}$ in \eqref{eq:ed} is exponential distributed  with PDF $f(\gamma^{(\rm d)}) =\frac{1}{\bar{\gamma}^{(\rm d)}}{\rm e}^{-\gamma^{(\rm  d)}/\bar{\gamma}^{(\rm d)}}$  where $\bar{\gamma}^{(\rm d)}$ is the average SNR. Using \eqref{eq:ed}, the expected energy for D2D relaying:
\begin{align}
\begin{split}
\bar{E}^{\rm D2D}&
=\left(\eta_1^{(\rm d)}+\eta_2^{(\rm d)}\mathbb{E}[P^{\rm ckt}]\right)\frac{1}{\bar{\gamma}^{(\rm d)}}\int_{\gamma_{\rm th}^{(\rm d)}}^{\infty} \frac{1}{\log_{e}(1+x)} e^{-x/\bar{\gamma}^{(\rm d)}}\mathrm dx
\end{split}
\label{eq:D2D_Energy}
\end{align}
where $\gamma_{\rm th}^{(\rm d)}$ is the threshold SNR in linear scale for D2D communication.
Using the series expansion of exponential function in \eqref{eq:D2D_Energy}, we get an exact expression of the expected energy consumption in D2D relaying
\begin{align}
\begin{split}
&\bar{E}^{\rm D2D}
=(\eta_1^{\rm d}+0.5\eta_2^{(\rm d)}(P_{\rm max}^{\rm ckt}+P_{\rm min}^{\rm ckt})) \times \sum_{k=0}^{\infty}\frac{(-1)^k}{k!}\frac{1}{(\bar{\gamma}^{(\rm d)})^{k+1}}[E_i(\gamma_{\rm max}+k\gamma_{\rm max})-E_i(\gamma_{\rm th}^{(\rm d)}+k\gamma_{\rm th}^{(\rm d)})].
\end{split}
\label{eq:d2d_series}
\end{align}

We also provide simple bounds on \eqref{eq:D2D_Energy} in the following Theorem:
\begin{my_theorem}
\label{theorem:d2d}
If  $P_{\rm min}^{\rm ckt}$ and $P_{\rm max}^{\rm ckt}$ are minimum and maximum circuit transmit   power of all devices, respectively, $\gamma_{\rm th}$ is the threshold SNR, and $\eta_1^{(\rm d)}=10\log_{e}(2)P^{(\rm d)}L/B$, $\eta_2^{(\rm d)}=\eta_1^{(\rm d)}/P^{(\rm d)}$,  then the expected energy consumption in D2D relaying under Rayleigh fading channel with average SNR $\bar{\gamma}^{(\rm d)}$ satisfies
\begin{align}
\begin{split}
&(\eta_1^{(\rm d)}+0.5\eta_2^{(\rm d)}(P_{\rm max}^{\rm ckt}+P_{\rm min}^{\rm ckt})) \Big(\frac{1}{\bar{\gamma}^{(\rm d)}}\log_e(1+\frac{\bar{\gamma}^{(\rm d)}}{\gamma_{\rm th}^{(\rm d)}})-\frac{1}{(\bar{\gamma}^{(\rm d)})^2}\log_e(1+\frac{\bar{\gamma}^{(\rm d)}}{\gamma_{\rm th}^{(\rm d)}})\Big)\leq\bar{E}^{\rm D2D}\\&\leq (\eta_1^{(\rm d)}+0.5\eta_2^{(\rm d)}(P_{\rm max}^{\rm ckt}+P_{\rm min}^{\rm ckt})) \Big(\frac{\bar{\gamma}^{(\rm d)}}{\bar{\gamma}^{(\rm d)}+\gamma_{\rm th}^{(\rm d)}}+\frac{1}{\bar{\gamma}^{(\rm d)}+\gamma_{\rm th}^{(\rm d)}}\log_e(1+\frac{\bar{\gamma}^{(\rm d)}}{\gamma_{\rm th}^{(\rm d)}})\Big)
\end{split}
\label{eq:theorem:d2d}
\end{align}
\end{my_theorem}
\begin{IEEEproof}
The	proof is given in Appendix 	E.
\end{IEEEproof}
From \eqref{eq:d2d_series} and  \eqref{eq:theorem:d2d}, it can be seen  that the expected energy decreases with an increase in the average SNR at the relaying device. Since the relay devices have high SNR due to close proximity with the source device in the D2D communication, the energy overhead of the relaying  among devices is  negligible as compared with the  transmission of data to the BS.

  \begin{figure}[t]
	\centering
	{\includegraphics[scale=0.54]{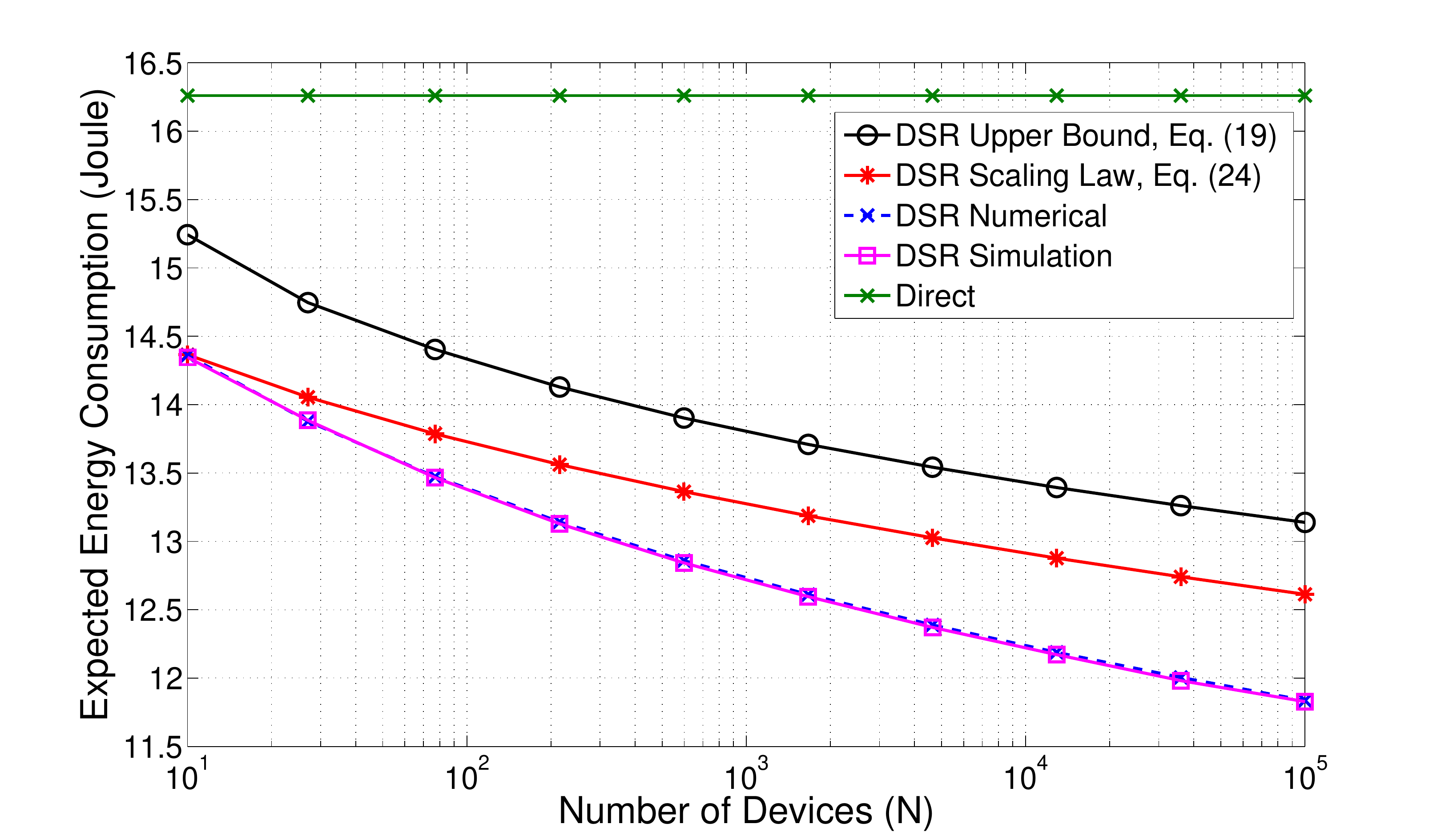}}
	\caption{Verification of scaling law and bounds.}
	\label{fig:bounds_scaling}
\end{figure}
\begin{table}[t]
	\renewcommand{\arraystretch}{1.5}
	\caption{Average energy consumption (in $\mu$J) of various overheads obtained using simulation under 3GPP model.}
	\label{table:overhead}
	\centering
	\begin{tabular}{c|c|c|c|c|c}
		$\bar{E}^{\rm RTR}_{\rm tx} $	&$\bar{E}^{\rm RTR}_{\rm rx} $& $\bar{E}^{\rm CTR}_{\rm tx} $ &$\bar{E}^{\rm CTR}_{\rm rx} $&	$\bar{E}^{\rm D2D}_{\rm tx} $ &		$\bar{E}^{\rm D2D}_{\rm rx} $\\
		\hline
		11.60&  4.50& 3.35 &  1.30&350.5& 135.4\\

	\end{tabular}
\end{table}

\section{Simulation and Numerical Analysis}
In this section, we  demonstrate the effectiveness of the proposed protocol through numerical analysis and simulations carried out in MATLAB.

First, we  verify the analytical bounds and scaling law derived in this paper by considering  a simplified transmission model and without  overhead energies, as depicted in Fig.~\ref{fig:bounds_scaling}. For each transmission, a packet length of $L=2$ \mbox{MB} is considered. We consider channel between devices to the BS to be log-normal distributed with a spreading factor of $4$ \mbox{dB} and a path loss exponent $\alpha =4$.   The channel between devices  is assumed to be Rayleigh fading with  a path loss exponent $\alpha =3$.	 The transmit power for each device is set to $23$ \mbox{dBm}. For scaling law verification,  we consider $M=4$, $c_M= 0. 99$, $\delta_M=\ln(N)$, $\delta_1=\delta_M/4$, $\delta_2=\delta_M/2$ and $\delta_3=3\delta_M/4$.   Fig.~\ref{fig:bounds_scaling} verifies the analytical bounds and the derived scaling law on the expected consumed energy in a network of $10$ to $10^5$ devices situated uniformly at a distance of $300$ \mbox{m} from the BS, situated in the center. 

Next, we demonstrate the proposed  schemes under a more realistic scenario using the 3GPP D2D channel model and simulation parameters in line with 3GPP recommendations \cite{TR36.843}. We emulate a single cell network with up to $150$ devices distributed uniformly in a radius of  $50$ \mbox{m} to $500$ \mbox{m} with a BS in the center. The background noise for each device and the BS is taken as $-174$ \mbox{dBm/HZ}. We consider $20$ \mbox{dB} of interference at the  BS.  We assume equal transmit power ($23$ \mbox{dBm})  and equal initial energy of  $0.72$mWh for all devices.  We use the energy model presented in \cite{lauridsen2014empirical} to compute the energy consumption of the devices for communication with the BS or D2D communication.  We assume that the  communication range for D2D relaying is $50$ \mbox{m}. The transmission bandwidth for each device is $200$ \mbox{KHz}.  The channel between the device  and the BS is urban macro log-normal shadowing (spreading factor $4$ \mbox{dB}) while the channel between devices is modeled as Rayleigh fading using Winner II\cite{winner2007d1}.   For each transmission, a data packet length of $L=1024$ bytes is considered, and the size of D2D request/reply data is $L^{(d)}=10$ bytes.    In contrast to the existing performance evaluations for relaying schemes, we also include the overhead energy into our simulation model.   

We present the components of average consumed energy for various overheads in Table \ref{table:overhead}. This is obtained using the realistic simulation in line with 3GPP recommendations and the 3GPP D2D channel model. 

We compare the proposed algorithms with the  opportunistic relaying (OR) selection criterion using the channel statistics for both hops \cite{Bletsas2006} and the single hop (devices to the BS) \cite{Krikidis2008}.  These opportunistic schemes are fundamental in wireless networks to harness the multi-user diversity.  For a  comparison, we use the same network scenario i.e., selection of devices within D2D range in contrast to the state of art procedure of selecting a relay device from the whole network. However, we also demonstrate (in Fig.~\ref{fig:d2d_nearby}) the advantage of selecting few nearby devices to achieve the near-optimal performance.

  \begin{figure}[t]
	\centering
	{\includegraphics[scale=0.54]{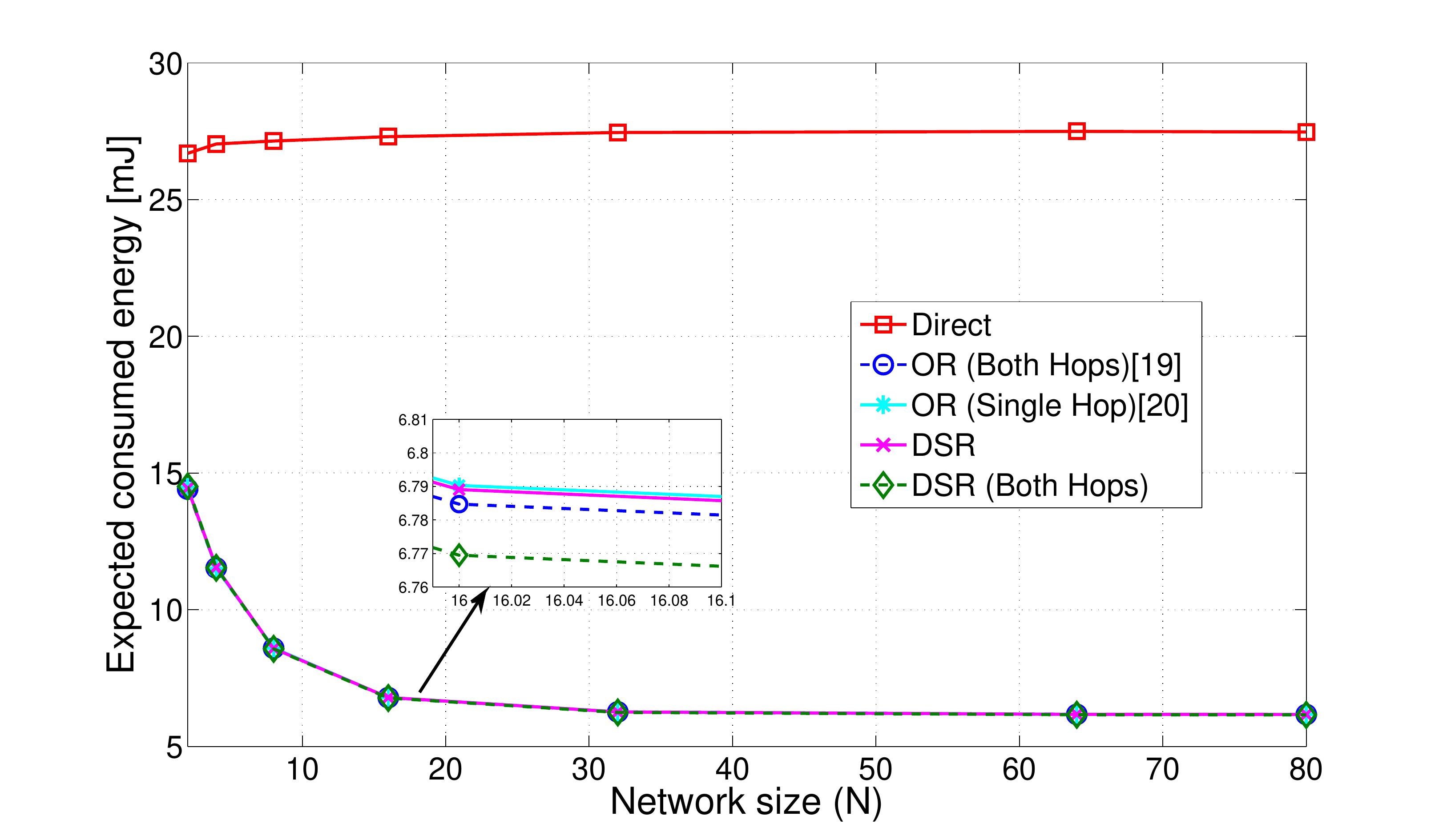}}
	\caption{Expected consumed energy by the D2D opportunistic relaying  schemes. }
	\label{fig:expected_energy}
\end{figure}

   \begin{figure}[t]
	\centering
	{\includegraphics[scale=0.54]{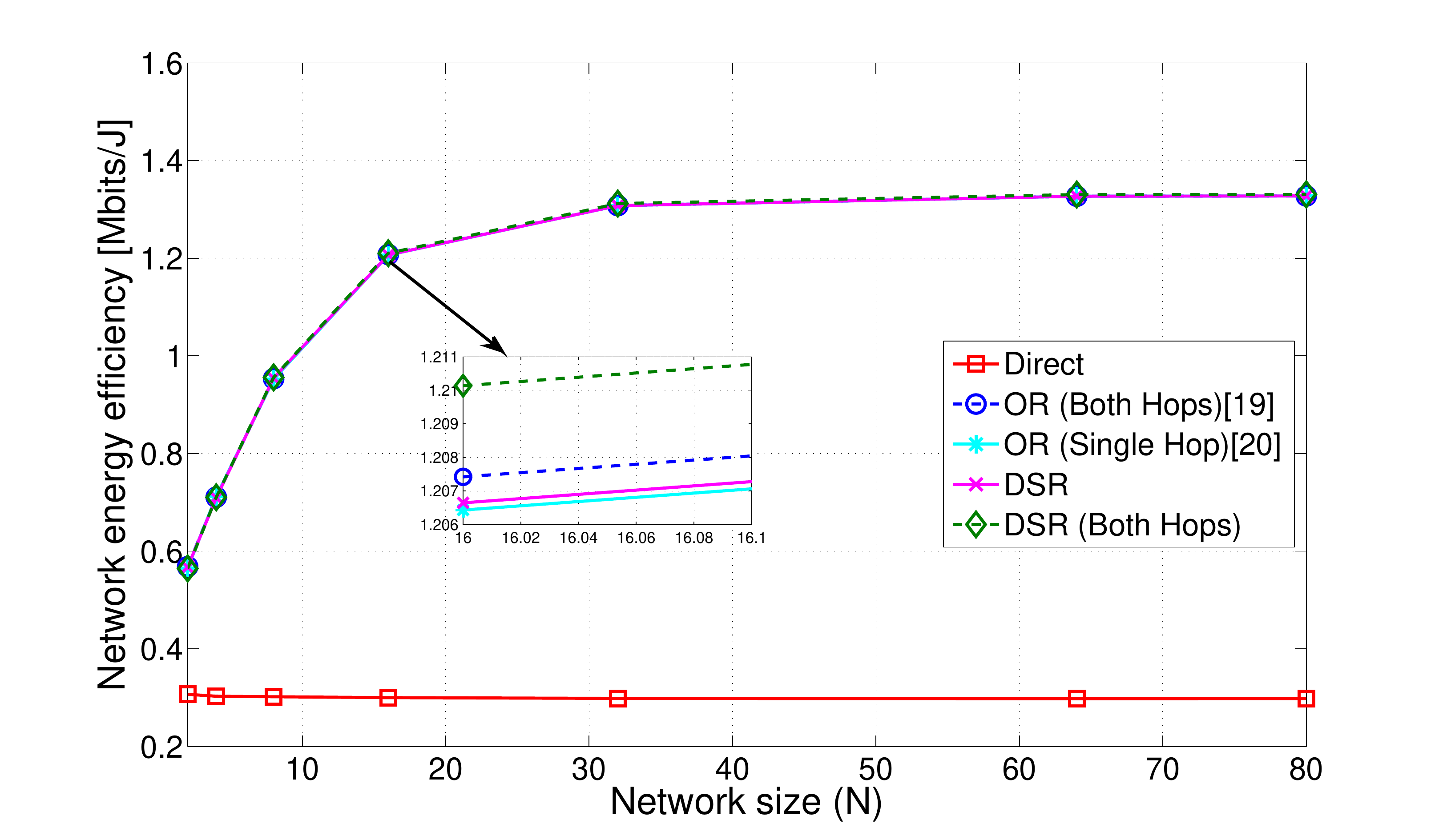}}
	\caption{Network energy efficiency of the D2D relaying  schemes. }
	\label{fig:energy_efficiency}
\end{figure}
\begin{figure}[t]
	\centering
	\label{rt}
	\includegraphics[scale=0.54]{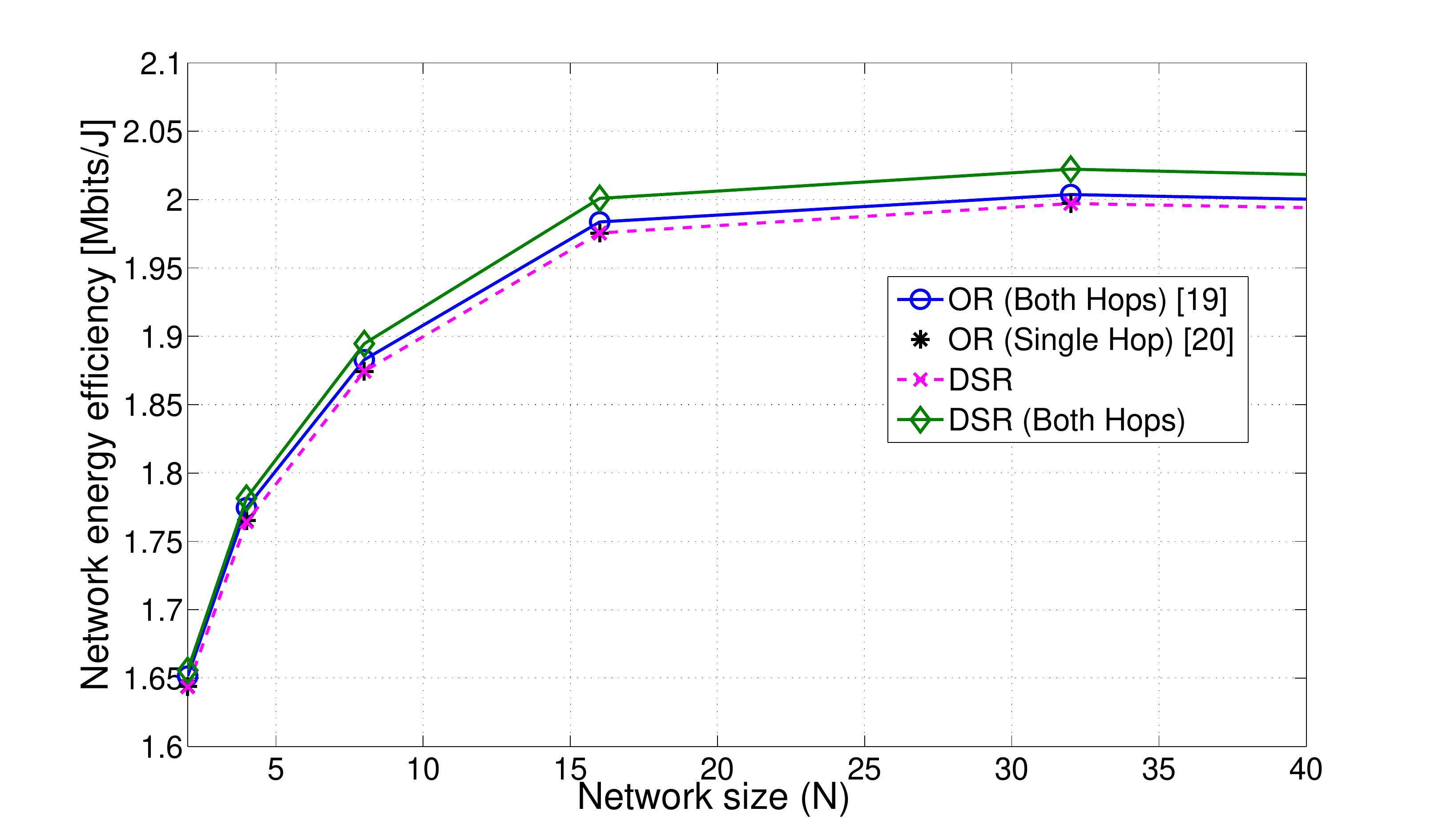}
	\caption{Energy efficiency performance of the relaying schemes by neglecting the interference at the BS. }
	\label{fig:d2d_ee_equal_noise}
\end{figure}

  \begin{figure}[t]
 	\centering
 	{\includegraphics[scale=0.54]{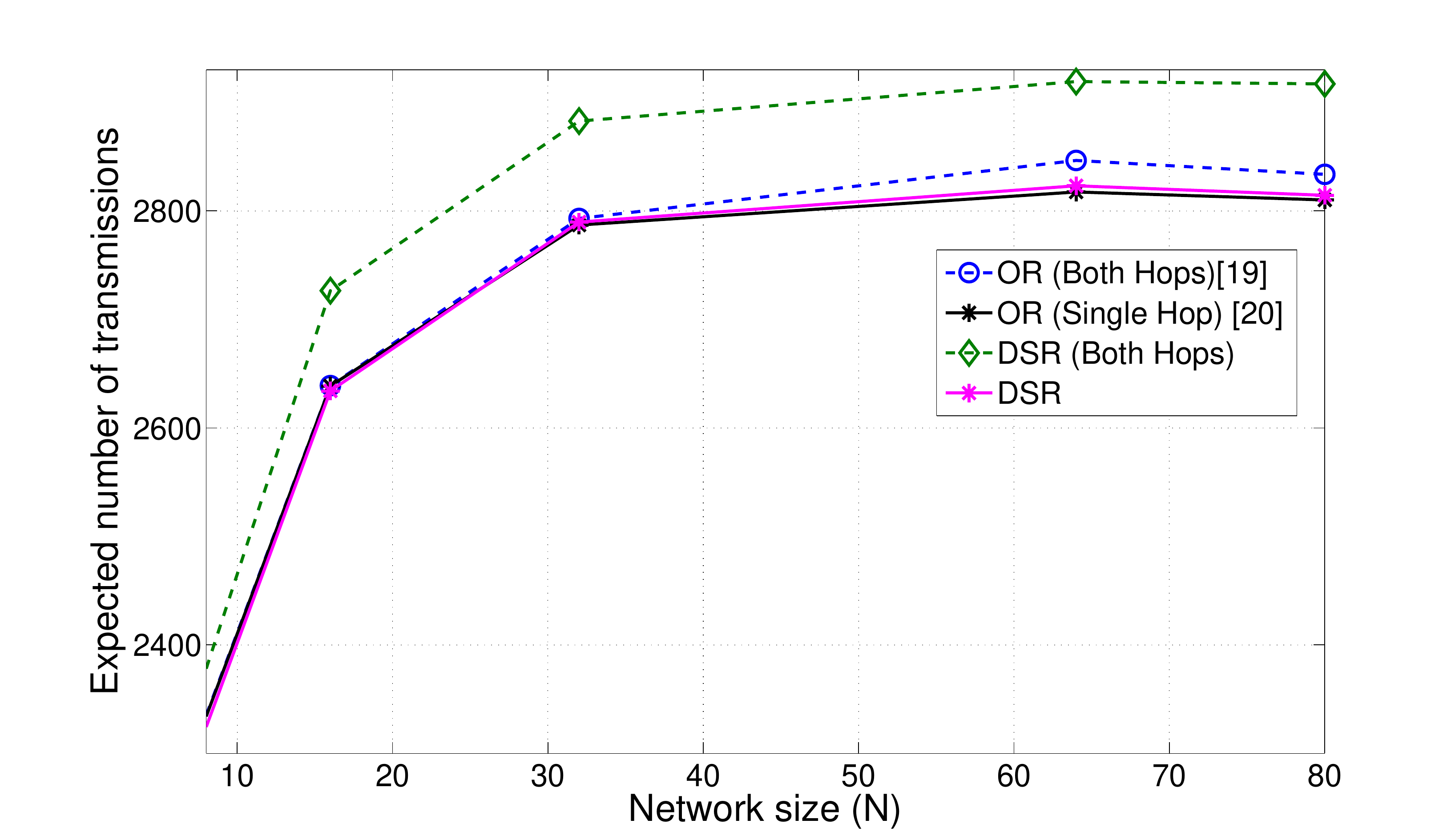}}
 	\caption{Expected number of transmissions of each devices until a device depletes its energy. }
 	\label{fig:perdevice_ntx}
 \end{figure}

\begin{figure}[t]
	\centering
	\label{rt}
	\includegraphics[scale=0.54]{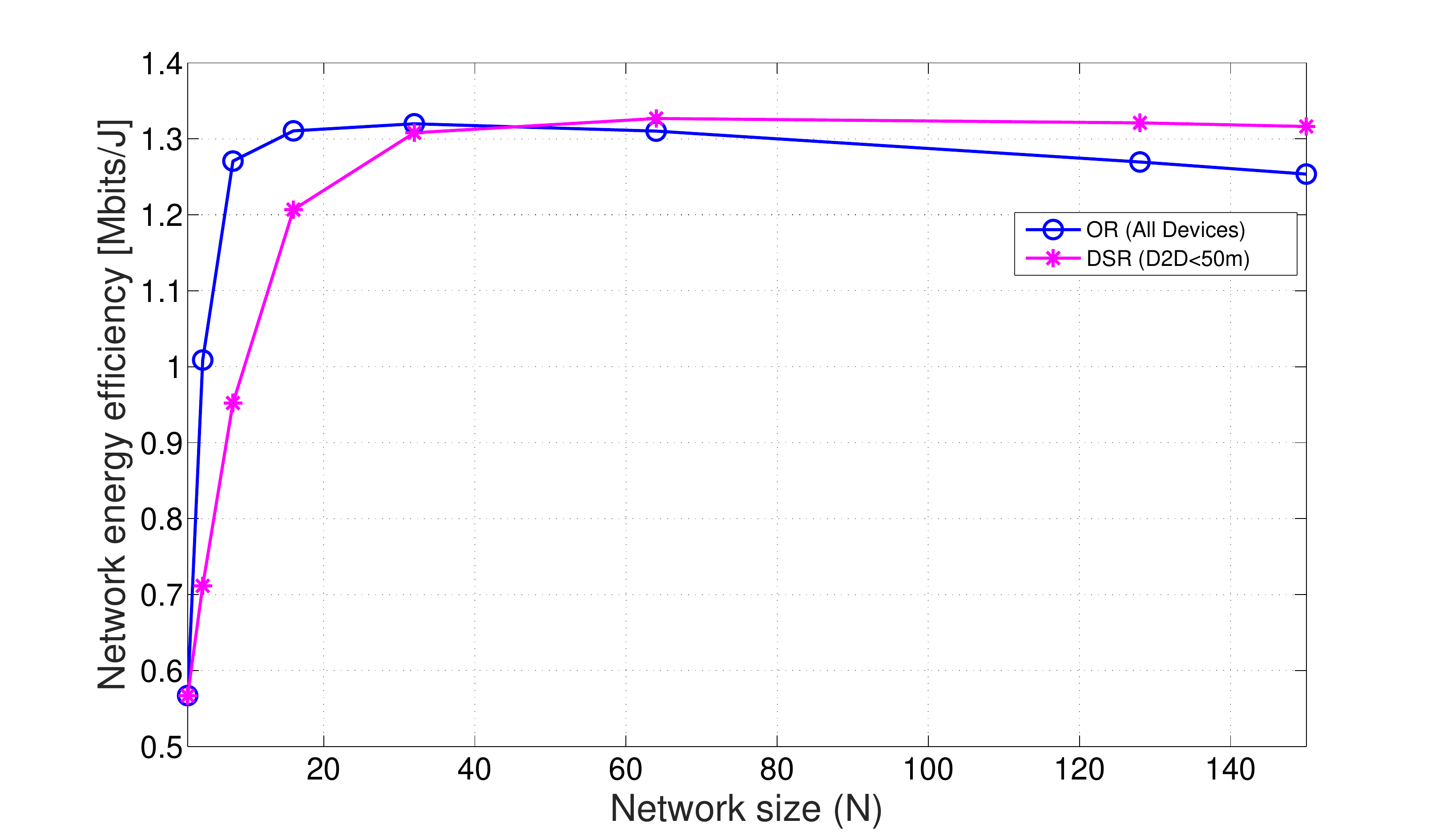}
	\caption{Performance comparison of single-hop DSR (relay devices under the range of $50$ \mbox{m} with the source) with the opportunistic scheme (single hop) based on relay selection from the whole network.}
	\label{fig:d2d_nearby}
\end{figure}


In  Fig.~\ref{fig:expected_energy}, we plot the expected consumed energy of the network by the  DSR protocols and compare the performance with SNR-based opportunistic schemes. The figure shows that the DSR based schemes provide significant performance improvement in terms of the expected consumed energy  compared to the direct communication with a few relaying devices i.e, within $N =15$, as shown in Fig.~\ref{fig:expected_energy}.  This happens because the log-normal shadowing of the second hop provides sufficient diversity to achieve the near-optimal performance with a few relaying devices.  Although the DSR selection criteria is based on the total consumed energy in contrast to the earlier SNR based OR scheme, the DSR achieves the similar performance to the OR since the impact of overhead energies and energy consumed in the circuitry is insignificant in the considered scenario. Fig.~ \ref{fig:expected_energy} also shows that the selection based on both hops performs very similar to the single hop schemes.

The energy efficiency of the network is depicted in Fig.~\ref{fig:energy_efficiency}. It can be seen that the network efficiency is increased by more than $4$ times  than the direct transmissions when $N>30$. As expected, the DSR achieves the similar  performance to the other opportunistic schemes. It can also be seen that the performance difference between the single-hop and the dual-hop is indistinguishable.  This happens because  the energy consumed by the devices in the second hop is dominant than the first hop (the second hop is limited by the log-normal shadowing and the BS is assumed to have an additional $20$ \mbox{dB} of  interference).   To this end, we compare the energy efficiency of the dual-hop schemes  with the single-hop  by neglecting the  $20$ \mbox{dB} of  interference at the BS, as shown in Fig.~\ref{fig:d2d_ee_equal_noise}. It can be seen that the DSR based on both hops achieves the best performance with an increase of $50$ Kbits/Joule comparing to the other schemes. 

Further,  we analyze the expected number of transmissions before the battery of the first device is depleted, as shown in Fig.~\ref{fig:perdevice_ntx}. It can be seen that the DSR (with both hops) outperform the other techniques, since it consumes the least energy for forwarding the data to the BS. The significant gain in the expected number of transmissions (in contrast to the other performance metrics such as expected energy consumption and energy efficiency) is due to the fact the small decrease in the transmission energy results into a significant cumulative gain in the number of transmissions by the dual-hop DSR when the first device depletes its energy.

Finally, we  demonstrate the use of D2D communication while selection of the relaying devices is limited to only nearby devices (considered in the range of  $50$ \mbox{m}). We show the performance for the DSR and compare it  with the existing methods of relay selection from the whole network, as depicted  in Fig.~\ref{fig:d2d_nearby}.  As demonstrated in Fig.~ \ref{fig:expected_energy} and Fig.~ \ref{fig:energy_efficiency}, the DSR achieves the near-optimal diversity performance using few devices only due to the log-normal shadowing. Thus, it is imperative to limit the relay selection to only few devices since it reduces latency  and minimizes the overhead required  for the selection of relay.    Fig.~\ref{fig:d2d_nearby} shows  that the performance of the DSR scheme is higher for large networks when  the relay is selected from only nearby devices (considered in the range of $50$ \mbox{m}). In case of  relay selection from whole network, the overhead in terms of energy increases due to a higher density of devices (more devices out of which the relay can be selected). It is noted that the performance degradation due to the overhead energy  depends on the size of RTS/CTS overhead message which is considered  $1$\% of the data length. Thus, in comparison to the existing protocol, the DSR has a higher performance gain in terms of  energy efficiency for large networks, as depicted in Fig.~\ref{fig:d2d_nearby}.

\section{Conclusion}
We have studied opportunistic relaying schemes for D2D communication and we have analyzed the energy consumed by the devices in a wireless network. We  have derived closed form expressions and analytical bounds on  D2D relaying protocols under log-normal shadowing  for both  direct and relayed transmissions. The analytical expressions show that the opportunistic schemes can achieve  significant performance gain  when the devices are in heavy shadowing area with respect to  the BS while the devices enjoy strong channel for inter-user D2D communication with negligible energy overhead.  Further, the derived scaling law on the consumed energy show that a near-optimal performance can be achieved in log-normal shadowing with a few devices only. This reduces the latency and overhead energy consumed by the devices in the selection  of relays.  We consider a  realistic cellular environment and  show that the opportunistic  D2D-relaying schemes improves the energy efficiency of the network comparing to the  direct communication using only few devices in the network. The results  will be  useful for millimeter-wave communications where non-line-of-sight communication  can be very hard.

\section*{Appendix A\\ Theorem 1:  Energy Consumption in Direct Transmission } 

 First we prove the upper bound. The integral in  \eqref{eq:E_direct1} can be represented as a sum of two integrals:
 \begin{align}
 \begin{split}
{\cal{I}}_{\rm ub}=\frac{1}{\sqrt{\pi} }\Big[\int_{0}^{\frac{\bar{\gamma}-\gamma_{\rm th}}{\sigma\sqrt{{2}}}}\frac{1}{\bar{\gamma}-\sqrt{2}t\sigma} e^{-t^2} \mathrm{d}{t} +\int_{0}^{\infty}\frac{1}{\bar{\gamma}+\sqrt{2}t\sigma} e^{-t^2} \mathrm{d}{t}\Big]
 \label{eq:lemma1_proof1}
 \end{split}
 \end{align}
Using $\exp[-x^2]\leq \frac{1}{1+x^2}$ in the first integral,  an upper bound on the integral
 ${\cal{I}}_1^{\rm DT}(\bar{\gamma},\sigma)$ can be obtained by the partial fraction method. This has been presented in 	\eqref{lemma1_direct_ub}.  Using standard mathematical procedure, an exact solution of the second integral ${\cal{I}}_2^{\rm DT}(\bar{\gamma},\sigma)$  is also presented in \eqref{lemma1_direct_ub}.  Using these, and average of uniform random variable, we get the upper bound \eqref{eq:theorem:dt} of Theorem \ref{theorem:dt}.

 For the lower bound, we use (\ref{eq:E_0}) and  $1+z\leq e^{z}$ to get the integral as
 \begin{align}
 \begin{split}
 {\cal{I}}_{\rm lb}=\frac{1}{\sqrt{2\pi} (\bar{\gamma}+1)}\int_{\frac{(\gamma_{\rm th}-\bar{\gamma}-1)}{\sigma}}^{\infty} e^{-\frac{x^2}{2}-\frac{\sigma}{\bar{\gamma}+1}x} \mathrm{d}{x}  
 \end{split}
 \label{eq:EE1}
 \end{align}
Completing the square in the exponential function and representing the integral into Gaussian Q-function with simple substitution,  we get the lower bound  \eqref{eq:theorem:dt} of Theorem \ref{theorem:dt}.
 
\section*{Appendix B\\ Theorem 2:  Energy Consumption in Relay Transmission}
First, we derive an  expression on energy consumed due the  circuit power using the order statistics on the uniform random variable in the following lemma.
\begin{my_lemma}
	\label{lemma:min_uniform}
Let ${Z_1, Z_2, \cdots Z_N}$ are $N$ i.i.d uniform random variables  in the interval $[a,b]$, and $Z_{(1)} = \min \{Z_1, Z_2, \cdots Z_N\}$.  The expected value  of  the minimum of a uniform random variable is
	\begin{align}
	\begin{split}
	\mathbb{E}[Z_{(1)}]= \frac{b+Na}{1+N}.
	\end{split}
	\label{eq:uniform_min}
	\end{align}
\end{my_lemma}
The proof follows standard procedure and has been omitted.

Using Lemma	\ref{lemma:min_uniform}, $\mathbb{E}[P^{\rm ckt}_{(1)}]$ can be derived. An upper bound on $I_{1}^{\rm RELAY}(N,\sigma)$ in \eqref{eq:dsr_main3} can be obtained using $Q(t)= 1-Q(-t)$ and Chernoff bound $Q(t)\leq \frac{1}{2}\exp[-t^2/2]$, and $\exp[-z]<\frac{1}{1+z}$ to convert into polynomial function:

\begin{align}
\begin{split}
I_{1}^{\rm RELAY}(N,\sigma)  \leq \frac{1 }{(2)^N}\int_{0}^{\frac{\bar{\gamma}-\gamma_{\rm th}}{\sigma}}\frac{1}{(\bar{\gamma}-t\sigma)^2(1+\frac{N}{2}t^2)}\mathrm {d} t
\end{split}
\label{eq:dsr_I1_1}
\end{align}

We use partial fraction to solve the integral in (\ref{eq:dsr_I1_1})  in an exact form, as given in (\ref{lemma2_I1_dsr_ub}).

To analyze  $I_{2}^{\rm RELAY}(N,\sigma)$,  we use the binomial expansion of $(1-Q(x))^N$ and interchange the summation and the integration to get 
\begin{align}
\begin{split}
&I_{2}^{\rm RELAY}(N,\sigma)= \sum_{k=0}^{N}{{N}\choose{k}}(-1)^k\int_{0}^{\infty}\frac{[Q(x)]^k}{(x\sigma+\bar{\gamma})^2}\mathrm {d} x
 =  \sum_{r=0}^{N}{{{N}/{2}}\choose{2r}}\int_{0}^{\infty}\frac{[Q(x)]^{2r}}{(x\sigma+\bar{\gamma})^2}\mathrm {d}x \\&
-\sum_{r=0}^{N}{{N/2}\choose{2r+1}}\int_{0}^{\infty}\frac{[Q(x)]^{2r+1}}{(x\sigma+\bar{\gamma})^2}\mathrm {d} x
\end{split}
\label{eq:dsr_12_binomial1}
\end{align}
Using standard mathematical procedure, we present a solution to the integral given in the following preposition.
\subsubsection*{Proposition B1}
	If $a, b, N>0$		
	\begin{align}
	\begin{split}
	&\int_{0}^{\infty}\frac{\exp[-Nx^2]}{(ax+b)^2}\mathrm {d} x=\Psi(N,a,b)  =\frac{1}{2a^3b}e^{-\frac{nb^2}{a^2}} \Bigg(2 \pi  b^2 N \text{erfi}\left(\frac{b
		\sqrt{N}}{a}\right)-2 b^2 N \text{Ei}\left(\frac{b^2 N}{a^2}\right)+2 a^2 e^{\frac{b^2 N}{a^2}}\\&-2 \sqrt{\pi } a b \sqrt{N} e^{\frac{b^2 N}{a^2}}-b^2 N \log \left(\frac{a^2}{b^2 N}\right)+b^2 N \log \left(\frac{b^2 N}{a^2}\right)+4 b^2 N \log \left(\frac{a}{b}\right)-2 b^2 N \log (N)\Bigg)
	\end{split}
	\label{eq:prep_integral}
	\end{align}

We use Chernoff  bounds: $f(\kappa)\exp[- \kappa x^2]\leq Q(x)\leq \frac{1}{2}\exp[-x^2/2]$ in 	(\ref{eq:dsr_12_binomial1}), and then use the result of (\ref{eq:prep_integral})	to get an upper bound on $I_{2}^{\rm RELAY}(N,\sigma)$ in	(\ref{lemma2_I2_dsr_ub}).

 Using Lemma \ref{lemma:min_uniform}, and  (\ref{lemma2_I1_dsr_ub}),  (\ref{lemma2_I2_dsr_ub})	 in \eqref{eq:dsr_main2} completes the proof the Theorem \ref{theorem:relay}.

\section*{Appendix C\\ Corollary 1:  Approximation on $I_{2}^{\rm RELAY}(N,\sigma)$}
	To derive an approximate expression on 	$I_{2}^{\rm RELAY}(N,\sigma)$, we use an approximation on  $ Q(x)\approx \exp[-(q_1x^2+q_2x+q_3)]$ and $e^{-z}\leq\frac{1}{1+z}, \forall z\leq0$  in (\ref{eq:dsr_12_binomial1}) to represent the integral 
	
	\begin{align}
	\begin{split}
	I_{2}^{\rm RELAY}(N,\sigma)\approx  \int_{0}^{\gamma_{\rm max}}\frac{\mathrm {d} x}{(x\sigma+\bar{\gamma})^2(1+k(q_1x^2+q_2x+q_3))}
	\end{split}
	\label{eq:appendix_lemma1_2}
	\end{align}
	where $\gamma_{\rm max}<\infty$ is chosen to avoid the divergence of the integral.  
	The integration in (\ref{eq:appendix_lemma1_2}) is derived in exact form as presented in 	(\ref{lemma2_I2_dsr_appr}). This completes the proof of Corollary \ref{corollary:I2}.

\section*{Appendix D\\ Theorem 3: Scaling Law on Energy Consumption}
We use $Q(0) =1/2$ to get an upper bound on the integral ${\cal{I}}_{1}^{\rm RELAY} (N,\sigma)$ in (\ref{eq:dsr_main3}):
\begin{align}
\begin{split}
{\cal{I}}_{1}^{\rm RELAY} (N,\sigma)\leq \frac{1}{2^N} (\frac{1}{\gamma_{\rm th}}-\frac{1}{\bar{\gamma}})
\end{split}
\label{eq:dsr_I1_scaling}
\end{align}
where the equality is achieved when $\gamma_{\rm th} =\bar{\gamma}$.
The integral ${\cal{I}}_{2}^{\rm RELAY} (N,\sigma)$ in (\ref{eq:dsr_main3}) can be decomposed:
\small
\begin{align}
\begin{split}
&{\cal{I}}_{2}^{\rm RELAY} (N,\sigma)= \int_{0}^{\delta_1}\frac{1}{(x\sigma+\bar{\gamma})^2}(1-Q(x))^N\mathrm {d} x
+\int_{\delta_1}^{\delta_2}\frac{1}{(x\sigma+\bar{\gamma})^2}(1-Q(x))^N\mathrm {d} x+\cdots +\int_{\delta_M}^{\infty}\frac{1}{(x\sigma+\bar{\gamma})^2}(1-Q(x))^N\mathrm {d} x
\end{split}
\label{eq:dsr_main4}
\end{align}
\normalsize
where $\delta_m>\delta_{m-1}>0$, $m=1,2,\cdots M$.
Since $Q(\delta_m)<Q(\delta_{m-1})$, we use the minimum of Q-function in each interval of integration to get an upper bound  (\ref{eq:dsr_main4}):
\small
\begin{align}
\begin{split}
&{\cal{I}}_{2}^{\rm RELAY} (N,\sigma)\leq\left(1-Q(\delta_1)\right)^N \frac{1}{\sigma}(\frac{1}{\bar{\gamma}}-\frac{1}{\sigma \delta_1+\bar{\gamma}})
+\left(1-Q(\delta_2)\right)^N \frac{1}{\sigma}(\frac{1}{\sigma \delta_1+\bar{\gamma}}-\frac{1}{\sigma \delta_2+\bar{\gamma}})+\cdots+\frac{1}{\sigma}(\frac{1}{\sigma \delta_M+\bar{\gamma}})
\end{split}
\label{eq:dsr_main9}
\end{align}
\normalsize
We use  $\delta_m= \sqrt{c_m\ln(N)}$ where $ 0\leq c_m\leq1$,  inequality $(1-x)^N\leq \frac{1}{1+Nx}$,  and a lower  bound on Q-function $Q(x)\geq \kappa e^{-x^2}$, where $\kappa=0.3885$ to bound $(1-Q(\delta_m))^N$: 
\begin{align}
\begin{split}
(1-Q(\delta_m))^N  \leq \frac{1}{1+\kappa N^{1-c_m}}
\end{split}
\label{eq:q_binomial}
\end{align}
Using \eqref{eq:q_binomial} in  \eqref{eq:dsr_main9}, we get 
\begin{align}
\begin{split}
&{\cal{I}}_{2}^{\rm RELAY} (N,\sigma)\leq \frac{1}{\sigma}\Big[\frac{1}{\bar{\gamma}+ \sigma\sqrt{c_M\ln(N)}}+\sum_{m=1}^{M-1}(\frac{1}{1+\kappa N^{(1-c_m)}}) (\frac{1}{\bar{\gamma}+\sigma \sqrt{c_{m-1}\ln(N)}}-\frac{1}{\bar{\gamma}+\sigma \sqrt{c_m\ln(N)}})\Big],
\end{split}
\label{eq:dsr_scaling_final}
\end{align}
where $c_0=0$.
Using Lemma \ref{lemma:min_uniform},  and \eqref{eq:dsr_I1_scaling},   \eqref{eq:dsr_scaling_final} in
\eqref{eq:dsr_main2}, and neglecting negative terms, we get \eqref{eq:theorem:scaling}. When $N\to \infty$, we get the scaling law for energy consumption of Theorem \ref{theorem:scaling}.
\section*{Appendix E\\ Theorem 4:  Energy Overhead of D2D Relaying}
Using \eqref{eq:uniform_min} and logarithm inequality $\frac{x}{x+1}\leq\log_{e}(1+x)\leq x $, the integral in \eqref{eq:D2D_Energy} for expected energy in D2D relaying can be represented in terms of exponential integral:
\begin{align}
\begin{split}
(\eta_1^{(\rm d)}+0.5\eta_2^{(\rm d)}(P_{\rm max}^{\rm ckt}+P_{\rm min}^{\rm ckt}))\frac{1}{\bar{\gamma}^{\rm d}}E_1(\frac{\gamma_{\rm th}^{(\rm d)}}{\bar{\gamma}^{d}})\leq\bar{E}^{\rm D2D}\leq\\ (\eta_1^{(\rm d)}+0.5\eta_2^{(\rm d)}(P_{\rm max}^{\rm ckt}+P_{\rm min}^{\rm ckt}))\Big(\exp(-\frac{\gamma_{\rm th}^{(\rm d)}}{\gamma^{\rm d}})+\frac{1}{\bar{\gamma}^{(\rm d)}}E_1(\frac{\gamma_{\rm th}^{(\rm d)}}{\bar{\gamma}^{(\rm d)}})\Big)
\end{split}
\end{align}

Further, we use the inequality on exponential integral 
$0.5\exp(-x)\log_{e}(1+2/x) <E_1(x) < \exp(-x)\log_{e}(1+1/x) $ and   $\exp(x)>1+x$ to get \eqref{eq:theorem:d2d} of Theorem \ref{theorem:d2d}.

\bibliographystyle{ieeetran}
\bibliography{d2d_bibtex}

 \end{document}